\documentclass[longauth]{aa}
\usepackage{graphicx}
\usepackage{natbib}
\usepackage[varg]{txfonts}
\usepackage[usenames,dvips]{color}

\usepackage{subfigure}
\usepackage{float}
\usepackage{xspace}
\usepackage{balance}
\usepackage{upgreek}
\usepackage{array}

\usepackage{hyperref}
\hypersetup{pagebackref=true, colorlinks=true,  linkcolor=red, citecolor=blue, urlcolor=blue, bookmarks=false}

\begin{document}

\title{The VIMOS Public Extragalactic Redshift Survey (VIPERS)
  \thanks{Based on observations collected at the European Southern
    Observatory, Cerro Paranal, Chile, using the Very Large Telescope
    under programmes 182.A-0886 and partly 070.A-9007.  Also based on
    observations obtained with MegaPrime/MegaCam, a joint project of
    CFHT and CEA/DAPNIA, at the Canada-France-Hawaii Telescope (CFHT),
    which is operated by the National Research Council (NRC) of
    Canada, the Institut National des Sciences de l'Univers of the
    Centre National de la Recherche Scientifique (CNRS) of France, and
    the University of Hawaii. This work is based in part on data
    products produced at TERAPIX and the Canadian Astronomy Data
    Centre as part of the Canada-France-Hawaii Telescope Legacy
    Survey, a collaborative project of NRC and CNRS.}}
  \subtitle{A precise measurement of the galaxy stellar mass function and the abundance of massive galaxies at redshifts $0.5<z<1.3$}

\author{
I.~Davidzon\inst{1,2}
\and M.~Bolzonella\inst{1}
\and J.~Coupon\inst{3}
\and O.~Ilbert\inst{4}
\and S.~Arnouts\inst{5,4}
\and S.~de la Torre\inst{6}
\and A.~Fritz\inst{7}
\and G.~De Lucia\inst{8}
\and A.~Iovino\inst{9}
\and B.~R.~Granett\inst{9}
\and G.~Zamorani\inst{1}
\and L.~Guzzo\inst{10,9}
\and U.~Abbas\inst{11}
\and C.~Adami\inst{4}
\and J.~Bel\inst{12}
\and D.~Bottini\inst{7}
\and E.~Branchini\inst{13,14,15}
\and A.~Cappi\inst{1,16}
\and O.~Cucciati\inst{1}           
\and P.~Franzetti\inst{7}
\and M.~Fumana\inst{7}
\and B.~Garilli\inst{7,4}     
\and J.~Krywult\inst{17}
\and V.~Le Brun\inst{4}
\and O.~Le F\`evre\inst{4}
\and D.~Maccagni\inst{7}
\and K.~Ma{\l}ek\inst{18}
\and F.~Marulli\inst{2,19,1}
\and H.~J.~McCracken\inst{20}
\and L.~Paioro\inst{7}
\and J.~A.~Peacock\inst{6}
\and M.~Polletta\inst{7}
\and A.~Pollo\inst{21,22}
\and H.~Schlagenhaufer\inst{23,24}
\and M.~Scodeggio\inst{7} 
\and L.~A.~M.~Tasca\inst{4}
\and R.~Tojeiro\inst{25}
\and D.~Vergani\inst{26}
\and A.~Zanichelli\inst{27}
\and A.~Burden\inst{25}
\and C.~Di Porto\inst{1}
\and A.~Marchetti\inst{28,9} 
\and C.~Marinoni\inst{12,29}
\and Y.~Mellier\inst{20}
\and L.~Moscardini\inst{2,19,1}
\and T.~Moutard\inst{4}
\and R.~C.~Nichol\inst{25}
\and W.~J.~Percival\inst{25}
\and S.~Phleps\inst{24}
\and M.~Wolk\inst{20}
}

   \offprints{\texttt{iary.davidzon@unibo.it}}  
\institute{
INAF - Osservatorio Astronomico di Bologna, via Ranzani 1, I-40127, Bologna, Italy 
\and Dipartimento di Fisica e Astronomia - Universit\`{a} di Bologna, viale Berti Pichat 6/2, I-40127 Bologna, Italy 
\and Institute of Astronomy and Astrophysics, Academia Sinica, P.O. Box 23-141, Taipei 10617, Taiwan 
\and Aix Marseille Universit\'e, CNRS, LAM (Laboratoire d'Astrophysique de Marseille) UMR 7326, 13388, Marseille, France  
\and Canada-France-Hawaii Telescope, 65--1238 Mamalahoa Highway, Kamuela, HI 96743, USA 
\and SUPA, Institute for Astronomy, University of Edinburgh, Royal Observatory, Blackford Hill, Edinburgh EH9 3HJ, UK 
\and INAF - Istituto di Astrofisica Spaziale e Fisica Cosmica Milano, via Bassini 15, 20133 Milano, Italy 
\and INAF - Osservatorio Astronomico di Trieste, via G. B. Tiepolo 11, 34143 Trieste, Italy 
\and INAF - Osservatorio Astronomico di Brera, Via Brera 28, 20122 Milano, via E. Bianchi 46, 23807 Merate, Italy 
\and Dipartimento di Fisica, Universit\`a di Milano-Bicocca, P.zza della Scienza 3, I-20126 Milano, Italy 
\and INAF - Osservatorio Astrofisico di Torino, 10025 Pino Torinese, Italy 
\and Aix-Marseille Universit\'e, CNRS, CPT (Centre de Physique  Th\'eorique) UMR 7332, F-13288 Marseille, France 
\and Dipartimento di Matematica e Fisica, Universit\`{a} degli Studi Roma Tre, via della Vasca Navale 84, 00146 Roma, Italy 
\and INFN, Sezione di Roma Tre, via della Vasca Navale 84, I-00146 Roma, Italy 
\and INAF - Osservatorio Astronomico di Roma, via Frascati 33, I-00040 Monte Porzio Catone (RM), Italy 
\and Laboratoire Lagrange, UMR7293, Universit\'e de Nice Sophia-Antipolis,  CNRS, Observatoire de la C\^ote d'Azur, 06300 Nice, France 
\and Institute of Physics, Jan Kochanowski University, ul. Swietokrzyska 15, 25-406 Kielce, Poland 
\and Department of Particle and Astrophysical Science, Nagoya University, Furo-cho, Chikusa-ku, 464-8602 Nagoya, Japan 
\and INFN, Sezione di Bologna, viale Berti Pichat 6/2, I-40127 Bologna, Italy 
\and Institute d'Astrophysique de Paris, UMR7095 CNRS, Universit\'{e} Pierre et Marie Curie, 98 bis Boulevard Arago, 75014 Paris, France 
\and Astronomical Observatory of the Jagiellonian University, Orla 171, 30-001 Cracow, Poland 
\and National Centre for Nuclear Research, ul. Hoza 69, 00-681 Warszawa, Poland 
\and Universit\"{a}tssternwarte M\"{u}nchen, Ludwig-Maximillians Universit\"{a}t, Scheinerstr. 1, D-81679 M\"{u}nchen, Germany 
\and Max-Planck-Institut f\"{u}r Extraterrestrische Physik, D-84571 Garching b. M\"{u}nchen, Germany 
\and Institute of Cosmology and Gravitation, Dennis Sciama Building, University of Portsmouth, Burnaby Road, Portsmouth, PO1 3FX 
\and INAF - Istituto di Astrofisica Spaziale e Fisica Cosmica Bologna, via Gobetti 101, I-40129 Bologna, Italy 
\and INAF - Istituto di Radioastronomia, via Gobetti 101, I-40129, Bologna, Italy 
\and Universit\`{a} degli Studi di Milano, via G. Celoria 16, 20130 Milano, Italy 
\and Institut Universitaire de France, 103, bd. Saint-Michel, F-75005 Paris, France 
}

\titlerunning{VIPERS galaxy stellar mass functions}
\authorrunning{I.~Davidzon et al.}

\date{Received March 19, 2013 - Accepted July 12, 2013}
%
%
%
\abstract {We measure the evolution of the galaxy stellar mass
  function from $z= 1.3$ to $z=0.5$ using the first $53\,608$
  redshifts of the ongoing VIMOS Public Extragalactic Survey (VIPERS).
  Thanks to its large volume and depth, VIPERS provides a detailed
  picture of the galaxy distribution at $z\simeq 0.8$, when the
  Universe was $\simeq 7$\,Gyr old.  We carefully estimate the
  uncertainties and systematic effects associated with the SED fitting
  procedure used to derive galaxy stellar masses.  We estimate the
  galaxy stellar mass function at several epochs between $z=0.5$ and
  $1.3$, discussing the amount of cosmic variance affecting our
  estimate in detail.  We find that Poisson noise and cosmic variance
  of the galaxy mass function in the VIPERS survey are comparable to
  the statistical uncertainties of large surveys in the local
  universe.  VIPERS data allow us to determine with unprecedented
  accuracy the high-mass tail of the galaxy stellar mass function,
  which includes a significant number of galaxies that are too rare to
  detect with any of the past spectroscopic surveys. At the epochs
  sampled by VIPERS, massive galaxies had already assembled most of
  their stellar mass.  We compare our results with both previous
  observations and theoretical models.  We apply a photometric
  classification in the $(U-V)$ rest-frame colour to compute the mass
  function of blue and red galaxies, finding evidence for the
  evolution of their contribution to the total number density budget:
  the transition mass above which red galaxies dominate is found to be
  about $10^{10.4}\,\mathcal{M}_\odot$ at $z \simeq 0.55,$ and it
  evolves proportionally to $(1+z)^3$.  We are able
  to separately trace the evolution of the number density of blue and
  red galaxies with masses above $10^{11.4}\,\mathcal{M}_\odot$, in a
  mass range barely studied in previous work.  We find that for such
  high masses, red galaxies show a milder evolution with redshift,
  when compared to objects at lower masses.  At the same time, we
  detect a population of similarly massive blue galaxies, which are no
  longer detectable below $z=0.7$. These results show the improved
  statistical power of VIPERS data, and give initial promising
  indications of mass-dependent quenching of galaxies at $z\simeq1$.}

%
\maketitle
\section{Introduction}
\label{Introduction}
The past decade has seen significant advances in the study of galaxy
evolution prompted by large astronomical surveys. In particular, such
surveys sample large cosmic volumes and collect large amounts of data,
thus facilitating a number of important statistical studies. The
galaxy stellar mass function (GSMF), defined as the co-moving number
density of galaxies within a stellar mass bin
$(\mathcal{M},\mathcal{M}\,+\,\mathrm{d}\mathcal{M})$, is one such
fundamental statistic, allowing the history of baryonic mass assembly
to be traced. Measurements of the GSMF help in constraining the cosmic
star formation rate \citep[SFR, e.g.][]{Behroozi2013} and in
investigating how galaxy properties change as a function of stellar
mass, redshift, and environments \citep[e.g.~in galaxy
clusters,][]{Vulcani2011}.

In the nearby universe, the GSMF has been measured to high accuracy by
exploiting the Two Micron All Sky Survey (2MASS), the 2dF Galaxy
Redshift Survey \citep[2dFGRS,][]{Cole2001}, and the Sloan Digital Sky
Survey \citep[SDSS, e.g.][]{York2000}. Its shape is parametrised well
by a double \citet{Schechter1976} function, with an upturn at
$\mathcal{M}\simeq 10^{10}\mathcal{M}_\odot$
\citep{Baldry2008,Li&White2009,Baldry2012}. Such bimodality, also
visible in the SDSS luminosity function \citep{Blanton2005b}, reflects
the existence of two distinct galaxy types: a population of
star-forming galaxies, with blue colours and disc-dominated or
irregular morphology, and a class of red early-type galaxies that, in
contrast, have their star formation substantially shut off
\citep{Kauffmann2003c,Franx2008,Bell2007}.

At higher redshift, such statistical studies are more challenging
because of the faintness of the objects. However, early seminal work
took advantage of the Hubble Space Telescope to construct samples of a
few hundred galaxies up to $z\simeq 3$, finding evidence of an
increase in the average stellar mass density with cosmic time
\citep{Rudnick2003,Dickinson2003,Fontana2003}.  Later, deeper surveys
were able to show the lack of evolution at the high-mass end of the
GSMF \citep[GOODS-MUSIC catalogue,][]{Fontana2006}, which contrasted
with an increase in galaxy density at lower masses \citep[VVDS
survey,][]{Pozzetti2007}. This is a result that is consolidated up to
$z\simeq 4$ by means of near- and mid-infrared data, which  
facilitate better estimates of the stellar masses
\citep{PerezGonzalez2008,Kajisawa2009}.  Although some disagreements
exist, such findings indicate that massive galaxies were assembled
earlier than those with lower stellar mass, suggesting that a
`downsizing in stellar mass' has taken place \citep{Fontanot2009}.

Besides these results, first attempts to study the GSMF by dividing
blue/active from red/quiescent objects provided interesting results,
despite the relatively limited statistics, and revealed that within
the GSMF the number of blue galaxies at intermediate masses (about
$10^{10}\,\mathcal{M}_\odot$) decreases as a function of cosmic time,
while the fraction of red galaxies increases
\citep{Bundy2006,Borch2006}.  This early work was extended using
larger galaxy samples \citep[as in COSMOS and
zCOSMOS,][]{Drory2009,Ilbert2010,Pozzetti2010} or very deep
observations \citep[GOODS-NICMOS survey,][]{Mortlock2011}, which
produced robust results for the evolution in number density of both
these galaxy populations. They also showed that a double Schechter
function is a good fit to the GSMF data out to $z\simeq 1$
\citep[][]{Pozzetti2010,Peng2010}.

A fundamental picture emerging from these studies is the
transformation of star-forming galaxies into ``red and dead'' objects
through some physical mechanism that halts the production of new
stars.  To distinguish between the various mechanisms proposed in the
literature \citep[e.g.][and reference therein]{Gabor2010}, it is
crucial to obtain precise and accurate measurements to constrain
theoretical models \citep{Lu2012,Mutch2013,Wang2013}.  Unfortunately,
such comparisons are hard, as on one side modelling galaxy evolution,
when based on $N$-body dark matter simulations 
\citep[e.g.][]{DeLucia&Blaizot2007,Bower2006,Guo2011,Guo2013},
requires a high level of complexity to parametrise all the physical
processes (star formation, supernova ejecta, etc.). On the
observational side, instead, it is hard to attain the precision
required to constrain models, especially for the most massive
galaxies, which are highly affected by sample variance and
small-number statistics. 
Moreover, uncertainties in redshift measurements and stellar mass
estimates make the analysis even more complicated
\citep{Marchesini2009,Marchesini2010}.

The latest galaxy surveys are helping with improved measurements of
the GSMF and could shed light on the discrepancies between data and
models \citep[BOSS,][]{Maraston2012}.  State-of-the-art analyses
provide new evidence suggesting the dependence on cosmic time and
stellar mass of the physical processes that extinguish star formation:
from $z=3$ to $z=1$, the density of quiescent galaxies increases
continuously for $\mathcal{M}\gtrsim 10^{10.8}\,\mathcal{M}_\odot$
\citep[][using UltraVISTA data]{Ilbert2013}, while at $z<1$ it evolves
significantly at lower masses (\citealp{Moustakas2013} using PRIMUS
data).  On the other hand, several issues remain open. In particular,
the role environment plays is still being debated
\citep[][]{Cucciati2010,Iovino2010,Bolzonella2010,Peng2010,Vulcani2013}.

Within this context, the VIMOS Public Extragalactic Redshift Survey
(VIPERS) provides a novel opportunity. As we describe here, this
survey provides a combination of wide angle coverage, depth, and
sampling that proves to be ideal for measuring the GMSF at $z\sim 1$
with unprecedented precision.  The large volume allows effective
probing of the massive end of the GSMF at these redshifts: at the
high-mass end, where a few interlopers can dramatically change the
shape of the GSMF, accurate spectroscopic redshift measurements are
crucial for avoiding contaminations.

In this paper we present the first measurements of the GSMF from the
up-to-date catalogue containing $\sim 55\,000$ objects; in this first
analysis we concentrate on the evolution of the GSMF from $z = 1.3$
down to $z = 0.5$, i.e.~within the range covered by the VIPERS data,
for the whole galaxy sample and separately for the blue and red
populations.  We also discuss in detail the sources of error and
potential systematic effects that could become dominant at the level
of precision on the GSMF allowed by the VIPERS data.

In Sect.~\ref{Data} we present the VIPERS galaxy catalogue
that has been used 
in this work, and describe how stellar masses have been estimated
through the SED fitting technique.  The global mass function is
presented in Sect.~\ref{From SM to GSMF}, along with a discussion on
the sample completeness and the main sources of uncertainties. We
compare those results with both previous surveys and models in
Sect.~\ref{Comparisons to previous work}.  In Sect.~\ref{Analysis by
  galaxy type}, after applying a colour classification, we study the
mass function (and the related number density) of red and blue
galaxies.  Our results are summarised in Sect.~\ref{Conclusions}.
Unless specified otherwise, our cosmological framework assumes
$\Omega_m=0.25$, $\Omega_\Lambda=0.75$, and $h_{70}=H_0/(70
\,\mathrm{km}\,\mathrm{s}^{-1}\,\mathrm{Mpc}^{-1})$.  All the
magnitudes are in the AB system \citep{Oke1974}.

\section{Data}
\label{Data}

\begin{figure*}
\includegraphics[width=0.99\textwidth]{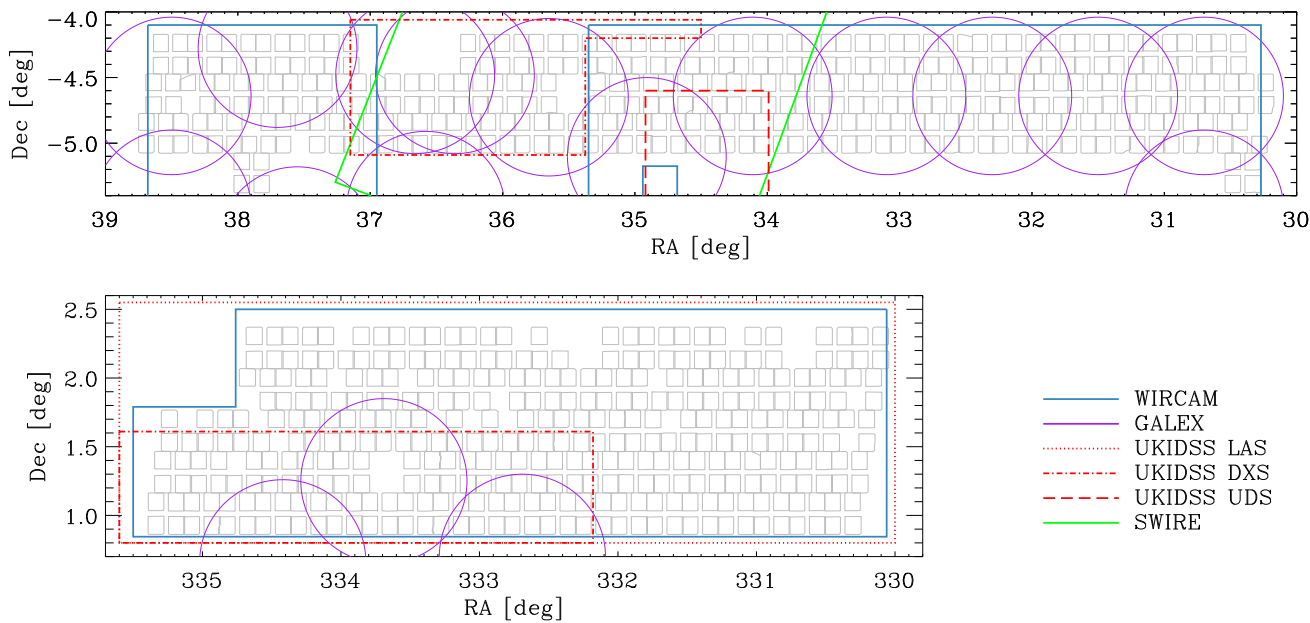} 
\caption{The coverage of ancillary data over the two VIPERS fields (W1
  and W4 in the upper and lower panels, respectively).  The W1 view is
  limited to the region sampled by VIPERS until now.  Each survey is
  shown with a different colour (see bottom right legend), while grey
  quadrants are the VIMOS pointings that led to the spectroscopic
  catalogue used in this work.}
\label{ancillar}
\end{figure*}

VIPERS\footnote{\url{http://vipers.inaf.it}} is an ongoing redshift
survey that aims at observing approximately $100\,000$ galaxies and
AGNs at intermediate redshifts ($\langle z\rangle \sim 0.8$) in the
magnitude range of $17.5\leqslant i \leqslant 22.5$.  At the
completion of the survey, expected in 2014, approximately
$24$\,deg$^2$ will have been covered within two fields of the
Canada-France-Hawaii Telescope Legacy Survey Wide
(CFHTLS-Wide)\footnote{\url{http://www.cfht.hawaii.edu/Science/CFHLS/}},
namely W1 and W4.  The sky region covered at present is $\sim
7.5\,{\rm deg}^2$ in each of them, with an effective area of
$5.34\,{\rm deg}^2$ in W1 and $4.97\,{\rm deg}^2$ in W4, after
accounting for the photometric and spectroscopic masks.  Once completed, VIPERS will be the
largest spectroscopic survey at such redshifts in terms of volume
explored ($1.5 \times 10^8\,\mathrm{Mpc}^3\,h_{70}^{-3}$).  All
details on the survey design and construction can be found in
\citet{Guzzo2013}.

The main science drivers of VIPERS are the accurate measurement of
galaxy clustering, bias parameter, and the growth rate of structures,
along with the study of the statistical properties of galaxies and
their evolution when the Universe was about half its current age.
These topics are the subject of the parallel accompanying papers of
this series
\citep{Guzzo2013,delaTorre2013,Marulli2013,Malek2013,Bel2013}.  A
previous smaller VIPERS sample has already been used to de-project
angular clustering in the CFHTLS full catalogue \citep{Granett2012}
and to develop a galaxy classification through principal component
analysis \citep{Marchetti2013}.

The spectroscopic survey is complemented by photometric ancillary data
(Fig.~\ref{ancillar}), obtained from public surveys and dedicated
observations, allowing us to estimate several galaxy properties with
high precision, in particular galaxy stellar masses and rest-frame
magnitudes.

\subsection{Photometry}
\label{Photometry}

The VIPERS spectroscopic sample has been selected from the W1 and W4
fields of the CFHTLS-Wide.  Therefore, for each galaxy we have a
photometric dataset consisting of $u^\ast$, $g^\prime$, $r^\prime$,
$i^\prime$, and $z^\prime$ magnitudes ({\tt SExtractor}'s MAG\_AUTO
derived in double image mode in order to maintain the same aperture in
all bands, \citealp{Bertin&Arnouts1996}), as measured by the Terapix
team for the T0005 data release \citep{Mellier2008}. The Terapix
photometric masks, which discard areas around bright stars or with
problematic observations, have been revisited by our team to recover
regions within those masks where the photometric quality is deemed
sufficient for our analysis \citep{Guzzo2013}.

We took advantage of the full wavelength range of the VIPERS
photometric dataset, since this significantly improves the results of
our SED fitting; in particular, near-infrared (NIR) fluxes are
critical to constraing physical parameters and break degeneracies
between the mean age of the stellar population and dust attenuation,
and they allow one to compute a robust estimate of stellar masses
\citep[e.g.][]{Lee2009}.

To exploit the full potential of VIPERS in analysing the galaxy
properties as a function of time and environment, we have undertaken a
follow-up in the $K$-band in the two VIPERS fields with the WIRCAM
instrument at CFHT and in the far- and near-UV ($FUV$ and $NUV$)
channel with the GALEX satellite (Arnouts et al., in prep.).  The
$K$-band observations were collected between 2010 and 2012 with
several discretionary time programmes.  The $K$-band depth has been
optimised to match the brightness of the spectroscopic sources: at the
magnitude limit ($K_{\rm WIRCAM}\simeq 22.0$ at $5\sigma$), $95$\% of
the spectroscopic sample in W4 is observed in $K_{\rm WIRCAM}$, while
in W1 this percentage is approximately $80$\% (see
Fig.~\ref{ancillar}).

In addition to WIRCAM data, we matched our CFHTLS optical catalogue
with the recent UKIDSS data releases\footnote{DR9 for LAS and DXS, DR8
  for UDS; \url{http://www.ukidss.org/}} using a matching radius of
$0.8\arcsec$.  The W1 field overlaps with UDS and DXS, whereas the W4
field is fully covered by the shallower LAS and partially covered by
DXS. Where available, we use Petrosian magnitudes in the $Y$, $J$,
$H,$ and $K$ bands converted in the AB system.  When also considering
$K_{\rm UKIDSS}$, the percentage of our spectroscopic sample with
$K$-band magnitude increases to $97$\% in W1 and $96$\% in W4.

We compared the $K$-band photometry for optical sources matched with
both UKIDSS and WIRCAM surveys, and find good agreement.  In fact, we
find a mean difference $\langle\Delta K\rangle = \langle
K_{\mathrm{WIRCAM}} - K_{\mathrm{UKIDSS}}\rangle \simeq -0.05$, with a
small dispersion $\sigma_{\Delta K}\simeq 0.10$ and $0.15$, for W1 and
W4, respectively.  These differences can be ascribed to the
transmission functions of the filters and the definition of the
aperture used when measuring magnitudes, and are close to photometric
errors.  To not overweight the $K$-band magnitudes in the SED fitting,
only the deeper $K_{\mathrm{WIRCAM}}$ data have been used when both
magnitudes were available for the same object.

The UV part of the spectrum can also be important for constraining the
galaxy dust content and the star formation rate.  We make use of
existing GALEX images observed with the deep imaging survey
(integration time $\sim 3 \times 10^4$\,s) in the $NUV$ and $FUV$
channels, and we have completed the coverage in W1 region with new
observations in the $NUV$ channel alone and with integration time
$T_{\rm exp}>1.5 \times 10^4$\,s.  Because of the GALEX large PSF
($\sim 5$\,arcsec), the source blending is a major issue in GALEX
deep-imaging mode.  To measure the UV fluxes of the sources, we use
the dedicated photometric algorithm EMphot \citep{Conseil2011}, which
adopts the positions of $U$-band selected priors and performs a
modelled PSF adjustment over small tiles based on the expectation
maximisation algorithm \citep{Guillaume2006}.  For our spectroscopic
sample, $63$\% ($15$\%) of the sources have an $NUV$ ($FUV$) flux
measurement in W1. In contrast, the W4 field has modest GALEX
coverage: $13$\% ($5$\%) of spectroscopic sources with an $NUV$
($FUV$) flux. The WIRCAM and GALEX datasets in the VIPERS fields are
described in Arnouts et al.~(in prep.).

Moreover, for $\sim 30$\% of the spectroscopic targets in W1, we also
took advantage of the SWIRE observations in the XMM-LSS field.  For
our SED fitting we only considered magnitudes in the $3.6\,\mu$m and
$4.5\,\mu$m bands, since beyond those wavelengths the survey is
shallower, and source detection is very sparse.  Moreover, at longer
wavelengths the re-emission from dust begins to contribute to the flux
of galaxies, and this feature is not reproduced by most of the models
of stellar population synthesis (see Sect.~\ref{Stellar masses}).

\subsection{Spectroscopy}
\label{Spectroscopy}

The spectroscopic catalogue used in this paper represents the first
$60$\% of VIPERS. This sample includes $53\,608$ galaxy spectra and
will be made available through the future VIPERS Public Data Release 1
(PDR-1). The VIPERS targets were selected via two criteria. The first
was aimed at separating galaxies and stars, and relies on the
combination of a point-like classification (based on measuring the
half-light radius) for the brightest sources and on comparing the five
optical magnitudes with galaxy and stellar spectral energy
distributions for the faintest ones \citep{Coupon2009}.  A fraction of
the point-like sources are targeted as AGN candidates, when located in
the AGN \textit{loci} of the two colour diagrams $(g-r)$ versus
$(u-g)$ and $(g-i)$ versus $(u-g)$. The second selection criterion,
based on $(g-r)$ and $(r-i)$ colours, was applied to exclude
low-redshift ($z < 0.5$) objects, and has been tested to ensure it
does not introduce any significant bias.  A complete description of
the whole source selection procedure is included in \citet{Guzzo2013}.

The spectroscopic observations were carried out using the VIMOS
instrument on VLT with the LR-Red grism ($R=210$), giving a wavelength
range of $5500$--$9500\,\mathrm{\AA}$ that guarantees the
observability of the main spectral features in the VIPERS redshift
range, e.g. the absorption lines CaII H \& K $\lambda\lambda
3934,3969$ and the emission line [OII] $\lambda 3727$.  Using a sample
of objects spectroscopically observed twice, we are able to estimate
an uncertainty of $\sigma_{z} = 0.00047 \, (1+z)$ for our measured
redshifts.

To maximise the multiplex capability of VIMOS, we adopted the
observational strategy described in \citet{Scodeggio2009msng} of using
shorter slits than in the previous surveys carried out with the same
instrument.  By virtue of this strategy, we reached a sampling rate of
approximately $40$\% with a single pass, essential to estimating the
large-scale environment (Cucciati et al., in prep.; Iovino et al., in
prep.).

The spectroscopic masks reproduce the footprint of the VIMOS
instrument, consisting of four quadrants and gaps between them for
each pointing, covering $224\,{\rm arcmin}^2$. Vignetted parts of the
quadrants have been removed to compute the effective area \citep[for a
detailed description see][]{Guzzo2013}.

Data reduction and redshift measurement were performed within the
software environment Easylife \citep{Garilli2012}, which is based on
the VIPGI pipeline \citep{Scodeggio2005} and EZ \citep[Easy
redshift]{Garilli2010}.  Once measured by the EZ pipeline and assigned
a confidence level, the spectroscopic redshifts were then checked and
validated independently by two team members. In case of any
discrepancy, they were reconciled by direct comparison.  In the vast
majority of cases, this involves spectra with very low signal-to-noise
ratios, which end up in the lowest quality classes.  In general, each
redshift is in fact assigned a confidence level, based on a
well-established scheme developed by previous surveys like VVDS
\citep{LeFevre2005} and zCOSMOS \citep{Lilly2009}. In detail, a
spectroscopic quality flag equal to $4$ corresponds to a confidence level of
$99.6$\%, with smaller flags corresponding to lower confidence levels,
as described in \citep{Guzzo2013}.  Objects with a single emission
line are labelled by flag $9$, and broad-line AGNs share the same
scheme, but their flags are increased by $10$.  Each spectroscopic
flag also has a decimal digit specifying the agreement with the
photometric redshift computed from CFHTLS photometry
\citep{Coupon2009}.

After excluding $3\,394$ galaxies with no redshift measurement (flag
$0$, which represents the lack of a reliable redshift estimate) and
$1\,750$ stars, our redshift sample contains $53\,608$ extragalactic
sources, nearly equally split between the two fields.  The quality of
redshift measurements for the sample with spectroscopic flags larger
than $2$, as estimated from the validation of multiple observations,
is high \citep[confidence $>95\%$, see][]{Guzzo2013}.

Since only a fraction of all the possible targets have been observed,
statistical weights are required to make this subsample representative
of all the galaxies at $i \leqslant 22.5$ in the survey volume.  Such
weights are calculated by considering the number of photometric
objects that have been targeted (target sampling rate, TSR), the
fraction of them classified as secure measurements (spectroscopic
success rate, SSR), and the completeness due to the colour selection
(colour sampling rate, CSR). The statistical weights can depend on the
magnitude, redshift, colour, and angular position of the considered
object. For each part of the statistical weight we considered only the
main and relevant dependencies, in order to avoid spurious
fluctuations when there are small subsamples. In particular, we
considered the TSR as a function of only the selection magnitude, the
SSR as a function of magnitude and redshift, and the CSR
\citep[estimated by using data from the VVDS flux limited
survey,][]{LeFevre2005} as a function of redshift.  Regarding the SSR,
only galaxies with quality flags between $2$ and $9$ ($\sim 41\,100$
galaxies in the redshift range $0.5 \leqslant z \leqslant 1.3$) were
considered in the analysis. (We exclude spectra classified as
broad-line AGNs.)  For a galaxy at redshift $z$ with magnitude $i$,
its statistical weight $w(i,z)$ is the inverse of the product of
TSR($i$), SSR($i,z$), and CSR($z$).  Once each galaxy in the
spectroscopic sample is properly weighted, we can recover the
properties of the photometric parent sample with good precision
\citep[for a detailed discussion on TSR, SSR, and CSR
see][]{Guzzo2013}.

\subsection{Stellar masses}
\label{Stellar masses}

\begin{figure}
\includegraphics[width=0.5\textwidth]{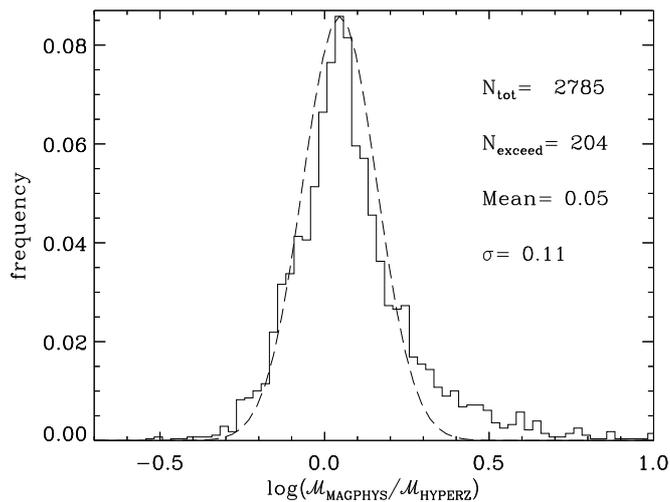}
\caption{Distribution of the differences between the values of stellar
  mass estimated using the two codes \textit{Hyperzmass} and
  \textit{MAGPHYS}.  Only results for the W1 field are shown (see
  text).  To limit the effects of parameter degeneracy, we restrict
  the comparison to galaxies that turn out to have solar metallicity,
  according to their best-fit templates both in \textit{Hyperzmass}
  and \textit{MAGPHYS}.  In this way the difference between
  $\mathcal{M}_\mathrm{MAGPHYS}$ and $\mathcal{M}_\mathrm{Hyperzmass}$
  cannot be due to a different metal content assumed in the two SED
  fitting estimates.  The dashed line gives the best-fitting Gaussian
  of the distribution, corresponding to the mean and standard
  deviation indicated.  Also indicated are the size of the galaxy
  subsample ($N_\mathrm{tot}$) and the number of stellar mass
  estimates for which the discrepancy is
  $\log(\mathcal{M}_\mathrm{MAGPHYS}/\mathcal{M}_\mathrm{Hyperzmass})
  > 2\sigma$ ($N_\mathrm{exceed}$).}
\label{2bursts} 
\end{figure}

Considering the small fraction of objects without $K$ band magnitude,
we decided to rely on SED fitting to derive stellar masses and to not
implement alternative methods, such as the \citet{Lin2007} relation
between stellar mass, redshift, and rest-frame magnitudes.

We thus derive galaxy stellar masses by means of an updated version of
\textit{Hyperzmass} \citep[][ software is available on
request]{Bolzonella2000,Bolzonella2010}.  Given a set of synthetic spectral
energy distributions, the software fits these models to the multi-band
photometry for each galaxy and selects the model that minimises the
$\chi^2$.  The SED templates adopted in this procedure are derived
from simple stellar populations (SSPs) modelled by Bruzual \& Charlot
\citep[hereafter BC03]{Bruzual2003}, adopting the \citet{Chabrier2003}
universal initial mass function (IMF)
\footnote{The choice of a different IMF turns into a systematic mean
  offset in the stellar mass distribution: for instance, our estimates
  can be converted to \citet{Salpeter1955} or \citet{Kroupa2001} IMF
  by a scaling factor of $\sim 1.7$ or $\sim 1.1$, respectively.}.
The BC03 model is one of the most commonly used ones
\citep[e.g.][]{Ilbert2010,Zahid2011,Barro2013}.  Another frequently
used SSP library is the one by \citet[][M05]{Maraston2005}, which
differs from the former because of the treatment of the thermally
pulsing asymptotic giant branch (TP-AGB) stellar phase, affecting NIR
emission of stellar populations aged $\sim 1\,{\rm Gyr}$.  The
question about the relevance of TP-AGB in the stellar population
synthesis is still open \citep[e.g.][]{Marigo2007}, with some evidence
that supports BC03 \citep{Kriek2010,Zibetti2013} in contrast to
observations favouring M05 \citep{MacArthur2010}.  In the following we
prefer to adopt the BC03 model, since most of the galaxies in the
redshift range we consider should not be dominated by the TP-AGB phase
\citep[which is instead relevant for galaxies at $1.4 \leqslant z
\leqslant 2.7$,][]{Maraston2006}.

The SSPs provided by \citet{Bruzual2003} assume a non-evolving stellar
metallicity $Z$, which we chose to be solar ($Z=Z_\odot$) or subsolar
($Z=0.2\,Z_\odot$).  This choice allows us to take the different
metallicities of the galaxies in our redshift range into account,
which can be lower than in the nearby universe \citep{Zahid2011},
without significantly increasing the effect of the age-metallicity
degeneracy.  Considering the low resolution of our spectroscopic
setup, it is difficult to put reliable constraints on $Z$ from the
observed spectral features, and therefore it was not possible to
constrain this parameter \textit{a priori}. Therefore, the metallicity
assigned to each galaxy is what is obtained from the best-fit model
(smallest $\chi^2$).

With respect to the galaxy dust content, we implemented the
\citet{Calzetti2000} and Pr\'evot-Bouchet
\citep{Prevot1984,Bouchet1985} extinction models, with values of $A_V$
ranging from $0$ (no dust) to $3$ magnitudes.  As pointed out in
previous work \citep[e.g.][]{Inoue2005,Caputi2008,Ilbert2009},
Calzetti's law is on average more suitable for the bluest SEDs, having
been calibrated on starburst (SB) galaxies, whereas the
Pr\'evot-Bouchet law is better for mild star-forming galaxies, since
it was derived from the dust attenuation of the Small Magellanic Cloud
(SMC) \citep[see also][]{Wuyts2011}.  Hereafter we refer to the
Calzetti and Pr\'evot-Bouchet models as SB and SMC extinction laws,
respectively.  We let the choice between the two extinction laws be
free, according to the best-fit model (smallest $\chi^2$), since we do
not have sufficient data at UV wavelengths to differentiate the
different trends of the two laws.

The SEDs constituting our template library are generated from the SSPs
following the evolution described by a given star formation history
(SFH). In this work, we assume exponentially declining SFHs, for which
$\mathrm{SFR}\propto\exp(-t/\tau)$, with the time scale $\tau$ ranging
from $0.1$ to $30\,\mathrm{Gyr}$.  A constant SFH (i.e., $\mathrm{SFR}
\sim 1\, \mathcal{M}_\odot \, \mathrm{yr}^{-1}$) is also considered.
This evolution follows unequally spaced time steps, from $t=0$ to
$t=20\,\mathrm{Gyr}$. No fixed redshift of formation is imposed in
this model.

Although such a parametrisation is widely used, recent studies have
shown how exponentially increasing SFHs can provide a more realistic
model for actively star-forming galaxies in which young stellar
populations outshine the older ones \citep{Maraston2010}.  This effect
becomes relevant at $z\sim 2$, when the cosmic star formation peaks,
and can be reduced by setting a lower limit on the age parameter, in
order to avoid unrealistic solutions that are too young and too dusty
\citep{Pforr2012}.  In our redshift range, galaxies whose SFH rises
progressively have low stellar masses
\citep[$\log(\mathcal{M}/\mathcal{M_\odot})\sim9.5$,][]{Pacifici2013}
falling below the limit of VIPERS. 
Moreover, \citet{Pacifici2013} identify a class of massive blue
galaxies that assembled their stellar mass over a relatively long
period, experiencing a progressive reduction of their star formation
at a later evolutionary stage.  For such \textit{bell-shaped} SFH,
neither increasing nor decreasing $\tau$-models seem to be suitable.
However, the resulting differences are smaller than the other
uncertainties of the SED fitting method \citep[cf.][]{Conroy2009}.

Another issue concerning the SFH is the assumption of smoothness.  In
fact, a galaxy could have experienced several phases of intense star
formation during its past, which can be taken into account by
superimposing random peaks on the exponential (or constant) SFR
\citep{Kauffmann2003c}.  Allowing the presence of recent secondary
bursts, thereby making the colours of an underlying old and red
population bluer, can lead to a systematically higher stellar mass
estimate.  However, only for a small fraction of objects is the
difference in $\mathcal{M}$ larger than $0.2$\,dex, as shown by
\citet{Pozzetti2007}.

We also quantified the effect of using complex SFHs in VIPERS, by
computing stellar masses using the \textit{MAGPHYS} package
\citep{daCunha2008}.  This code parametrises the star formation
activity of each galaxy template starting from the same SSP models as
\textit{Hyperzmass} (i.e.,~BC03), but using two components in the SFH,
namely an exponentially declining SFR and a second component of
additional bursts randomly superimposed on the former according to
\citet{Kauffmann2003c}.  The probability of a secondary burst
occurring is such that half of the galaxy templates in the library
have experienced a burst in their last $2\,\mathrm{Gyr}$. Each of
those episodes can last $3\times10^7$--$3\times10^8\,$yr, producing
stars at a constant rate.  The ratio between the stellar mass produced
in a single burst and the one formed over the entire galaxy's life by
the underlying exponentially declining model is distributed
logarithmically between $0.03$ and $4.0$.  The dust absorption model
adopted in \textit{MAGPHYS} is the one proposed by
\citet{Charlot&Fall2000}, which considers the optical depth of H II
and H I regions embedding young stars along with the extinction caused
by diffuse interstellar medium.  \textit{MAGPHYS} treats attenuation
in a consistent way, including dust re-emission at infrared
wavelengths; however, this feature does not represent a significant
advantage when dealing with VIPERS data since infrared magnitudes are
too sparse in our catalogue.  Metallicity values are distributed
uniformly between $0.02$ and $2\, Z_\odot$.  The wide range of tightly
sampled metallicities, the different model for the dust extinction,
and in particular the complex SFHs in the \textit{MAGPHYS} library are
the major differences with respect to the \textit{Hyperzmass} code.

In Fig.~\ref{2bursts} we compare the estimates obtained through
\textit{MAGPHYS} and \textit{Hyperzmass}, and verify that complex SFHs
have a minimal impact on the results (see Sect.~\ref{Other
  uncertainties}). Since \textit{MAGPHYS} requires a much longer
computational time than other SED fitting codes, we only estimate the
stellar mass for galaxies in the W1 field between $z=0.5$ and $z=1.3$.
Moreover, for this comparison we selected objects with the same
(solar) metallicity in both the SED fitting procedures, because in
this way we are able to investigate the bias mainly thanks to the
different SFH parametrisations. The distribution of the ratio between
the two mass estimates is reproduced well by a Gaussian function plus
a small tail towards positive values of
$\log(\mathcal{M}_\mathrm{MAGPHYS}/\mathcal{M}_\mathrm{Hyperzmass})$.
We find a small offset ($\langle\Delta\log\mathcal{M} \rangle =\langle
\log(\mathcal{M}_\mathrm{MAGPHYS}/\mathcal{M}_\mathrm{Hyperzmass})
\rangle \simeq 0.05$) and a small dispersion
($\sigma_{\Delta\mathcal{M}} \simeq 0.11$) for most of the galaxy
population, with significant differences between \textit{MAGPHYS} and
\textit{Hyperzmass} (i.e., $\Delta\log\mathcal{M} > 0.22$) for only
$\sim 7$\% of the testing sample ($N_\mathrm{exceed}$ in
Fig.~\ref{2bursts}).  The consequences on the GSMF are discussed in
Sect.~\ref{Other uncertainties}.

Given the wide range of physical properties allowed in the SED fitting
procedure, we decided to exclude some unphysical parameter
combinations from the fitting.  In particular, we limit the amount of
dust in passive galaxies (i.e., we impose $A_V \leqslant 0.6$ for
galaxies with ${\rm age}/\tau > 4$), we avoid very young extremely
star-forming galaxies with short $\tau$ timescales (i.e. we prevent
fits with models with $\tau \leqslant 0.6\,{\rm Gyr}$ when requiring
$z_{\rm form}< 1$), and we only allow ages to be within $0.1$\,Gyr and
the age of the Universe at the spectroscopic redshift of the fitted
galaxy \citep[see][]{Pozzetti2007,Bolzonella2010}.

According to \citet{Conroy2009}, the uncertainties associated with the
SED fitting can be $\sim 0.3$\,dex when considering all the possible
parameters involved and their allowed ranges.  In particular, given
the non-uniform coverage of the GALEX and SWIRE ancillary data matched
with our sample, we checked that the variation in the magnitude set
from one object to another does not introduce significant bias.  For
the subsample of galaxies with $FUV$, $NUV$, $3.6\,\mu$m, and
$4.5\,\mu$m bands available, we also estimate the stellar mass using
just the optical-NIR photometry. We find no systematic difference in
the two estimates of stellar mass (with and without the UV and
infrared photometry) and only a small dispersion of about $0.08$\,dex.

In summary, the VIPERS galaxy stellar mass estimates are obtained
using the BC03 population synthesis models with Chabrier IMF, smooth
(exponentially declining or constant) SFHs, solar and subsolar
metallicity, and the SB and SMC laws for modelling dust extinction.
Unless stated otherwise, this is the default parametrisation used
throughout this paper.

\section{From stellar masses to the galaxy stellar mass function}
\label{From SM to GSMF}

In this section we exploit the VIPERS dataset described above by
considering only our fiducial sample of $41\,094$ galaxies at
$z=[0.5,1.3]$ with spectroscopic redshift reliability $> 95$\% (see
Sect.~\ref{Spectroscopy}). As mentioned above, broad-line AGNs ($\sim
850$ in the present spectroscopic sample) are naturally excluded from
the sample, being visually identified during the redshift measurement
process. Instead, narrow-line AGNs are not removed from our sample,
but they do not constitute a problem for the SED fitting derived
properties, since in most of the cases their optical and NIR emission
are dominated by the host galaxy \citep{Pozzi2007}.  First of all, we
try to identify the threshold above which the sample is 
complete, and therefore the mass function can be considered reliable.  After that, we
derive the GSMF of VIPERS in various redshift bins and discuss the
main sources of uncertainty affecting it.

\subsection{Completeness}
\label{Completeness}

\begin{figure}
\centering
\includegraphics[width=0.48\textwidth]{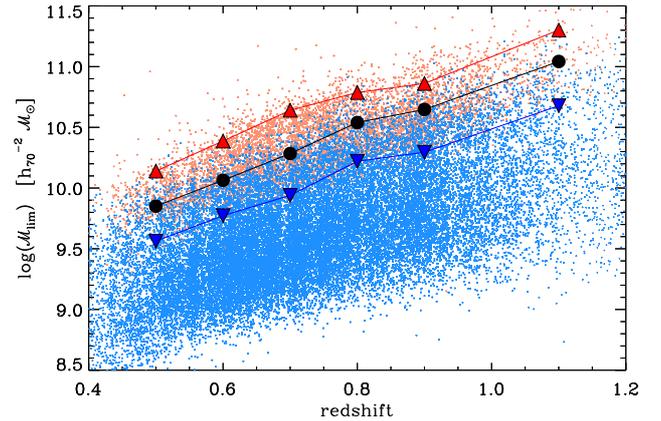}
\caption{The mass completeness threshold $\mathcal{M}_\mathrm{lim}$ as
  a function of redshift, computed for the total sample (the one used
  in Sect.~\ref{Global GSMF}, filled circles) and for the red (upward
  triangles) and blue (downward triangles) populations, defined as
  discussed in Sect.~\ref{Analysis by galaxy type}. In each redshift
  bin, the $\mathcal{M}_\mathrm{lim}$ estimate relies on the rescaled
  stellar mass $\mathcal{M}(i=i_\mathrm{lim})$ of the $20$\% faintest
  galaxies (see text).  We show $\mathcal{M}(i=i_\mathrm{lim})$ of the
  red and blue galaxies with small dots of analogous colours.}
\label{Mlim}
\end{figure}

In the literature, the completeness mass limit of a sample at a given
redshift is often defined as the highest stellar mass a galaxy could
have, when its observed magnitude matches the flux limit
\citep[e.g.][]{PerezGonzalez2008}.  This maximum is usually reached by
the rescaled SED of an old passive galaxy.  However, this kind of
estimate gives rise to a threshold that tends to be too conservative.
The sample incompleteness is due to galaxies that can be potentially
missed, because their flux is close to the limit of the survey.
Depending on the redshift, such a limit in apparent magnitude can
correspond to faint luminosities; in that case, only a small fraction
of objects will have a high stellar mass-to-light ratio, since blue
galaxies (with lower $\mathcal{M}/L$) will be the dominant population
\citep[e.g.][]{Zucca2006}. Thus, if based on the SED of an old passive
galaxy, the determination of the stellar mass completeness is somehow
biased in a redshift range that depends on the survey depth \citep[see
also the discussion in][Appendix C]{Marchesini2009}.

To avoid this problem, we apply the technique devised by
\citet{Pozzetti2010}. This procedure yields, for a given redshift and
flux limit, an estimate of the threshold $\mathcal{M}_\mathrm{lim}$
below which some galaxy type cannot be detected any longer.  Following
this approach, we estimate the stellar mass each object would have if
its magnitude, at the observed redshift, were equal to the $i$-band
limiting magnitude $i_\mathrm{lim}$.  This boundary mass
$\mathcal{M}(i\!=\!i_\mathrm{lim})$ is obtained by rescaling the
original stellar mass of the source at its redshift,
i.e.~$\log\mathcal{M}(i\!=\!i_\mathrm{lim})=\log\mathcal{M} +
0.4(i-i_\mathrm{lim})$.  The threshold $\mathcal{M}_\mathrm{lim}$ is
then defined as the value above which $90$\% of the
$\mathcal{M}(i\!=\!i_\mathrm{lim})$ distribution lies.  According to
this, at values higher than $\mathcal{M}_\mathrm{lim}$, our GSMF can
be considered complete.  We include in the computation only the $20$\%
faintest objects to mitigate the contribution of bright red galaxies
with large $\mathcal{M}/L$ when they are not the dominant population
around the flux limit, as they may cause the bias discussed at the
beginning of this section.

Since the $1/V_\mathrm{max}$ method \citep[][see Sect.~\ref{Global
  GSMF}]{Schmidt1968} intrinsically corrects the sample incompleteness
above the lower limit of the considered redshift bin
($z_\mathrm{inf}$), we apply to each redshift bin the
$\mathcal{M}_\mathrm{lim}$ computed by considering the objects inside
a narrow redshift interval $\Delta z = 0.05$ centred on
$z_\mathrm{inf}$.  Figure~\ref{Mlim} shows $\mathcal{M}_\mathrm{lim}$
as a function of redshift for the global and for the red and blue
samples used in Sect.~\ref{Analysis by galaxy type}, as well as the
value of $\mathcal{M}(i\!=\!i_\mathrm{lim})$ for each red and blue
galaxy.  As expected, the limiting mass increases as a function of $z$
and the values for red galaxies are significantly higher
($\sim0.5\,\mathrm{dex}$) than for the blue ones.

\begin{figure}[b!]
\centering
\includegraphics[width=0.48\textwidth]{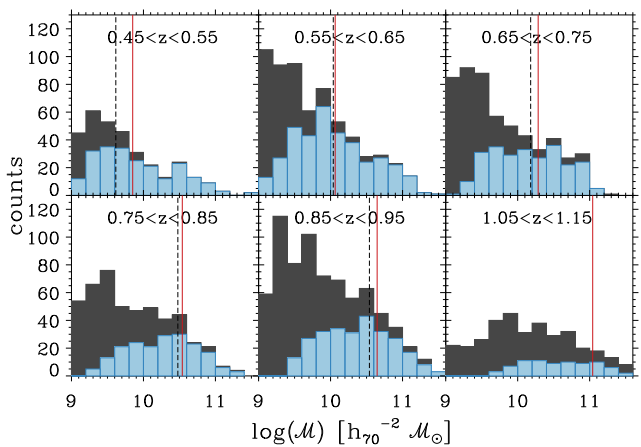}
\caption{Distributions of stellar masses in six redshift bins for the
  VVDS-Deep sample in the CFHTLS-W1 field at its limiting magnitude
  ($I \leqslant 24$, dark histograms), compared to the subset obtained
  by applying a magnitude cut similar to VIPERS, at $I \leqslant 22.5$
  (blue histograms).  In each panel, the black dashed line represents
  the limiting mass for the VVDS sample with $I \leqslant 22.5$. The
  red solid line instead gives the limiting mass for the VIPERS sample
  in the W1 field. Both limits, in good agreement with each other,
  correctly identify the threshold below which the shallower sample
  starts to miss a significant fraction ($> 20\%$) of objects.}
\label{VVDStest}
\end{figure}

In the context of the zCOSMOS project \citep{Lilly2009}, the approach
of \citet{Pozzetti2010} produced completeness limits in good agreement
with those obtained through mock survey samples \citep{Meneux2009}.
In VIPERS, we successfully tested our $\mathcal{M}_\mathrm{lim}$
estimates by taking advantage of the VVDS-Deep field, which is located
in the W1 field \citep[see][Fig.~2]{Guzzo2013}.  The VVDS sample
provides us with spectroscopically observed galaxies down to a fainter
limit, i.e.~$I_{\rm AB}=24$ \citep{LeFevre2005}.  Since the CFHTLS-W1
field contains both VVDS and part of VIPERS, we can compare the
stellar masses by relying on a similar photometric baseline
$(u,g,r,I,i,z,J^*,K^*)$.  When applying a VIPERS-like magnitude cut
($I<22.5$), we can find the fraction of missed objects with respect to
the parent $I<24$ sample as a function of stellar mass.  This test is
shown in Fig.~\ref{VVDStest}, where we compare the
$\mathcal{M}_\mathrm{lim}$ values of VVDS (limited to $I \leqslant
22.5$) and VIPERS to the distribution of stellar masses belonging to
the deeper (i.e., $I\leqslant24$) VVDS sample.  The
$\mathcal{M}_\mathrm{lim}$ values we computed \textcolor{red}{are close} to the threshold\textcolor{red}{s}
at which the stellar mass distribution starts to be incomplete with
respect to the deep VVDS sample (i.e.~the limit where the $I<22.5$
sample recovers less than $80$\% of the parent sample).

\subsection{Evolution of the mass function for the global population}
\label{Global GSMF}

\begin{figure*}
\centering
\includegraphics[width=0.95\textwidth]{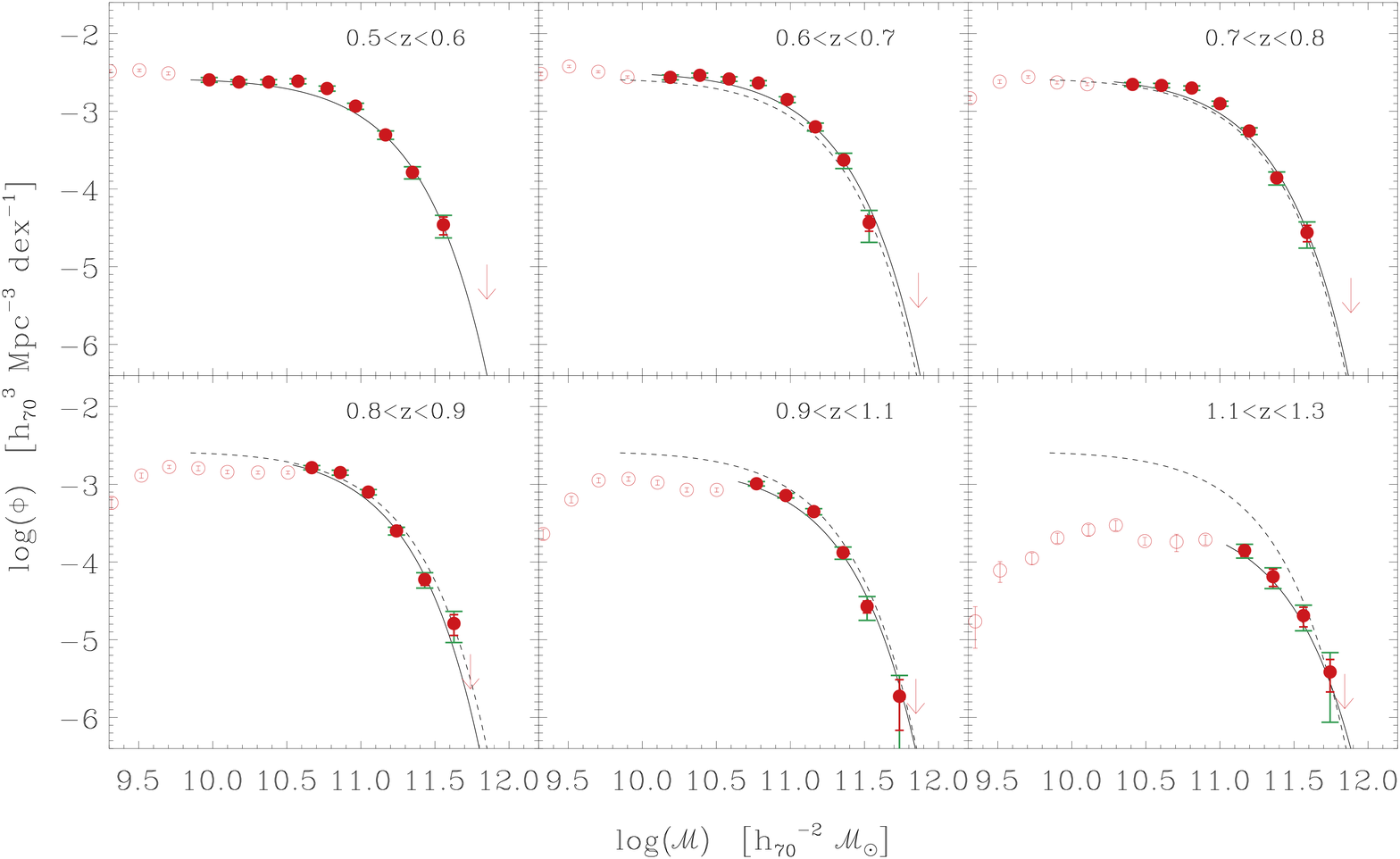}
\caption{The VIPERS galaxy stellar mass function at different
  redshifts.  Circles give the values determined through
  $1/V_\mathrm{max}$ in mass bins of $\Delta\mathcal{M}=0.2$\,dex; the
  centre of each bin corresponds to the weighted mean mass of the
  objects within it. Empty and filled symbols correspond to values
  below and above the completeness limit, respectively (see
  Sect.~\ref{Completeness}).  For the latter points, the red error
  bars show the uncertainty due to Poisson noise, while green bars
  account for Poisson noise and cosmic variance.  In each panel, a
  solid line shows the Schechter best-fit to the GSMF filled points,
  with the dashed line reproducing that of the first redshift bin, as
  a reference. The downward arrows give an upper limit to $\Phi$ where
  no detection is available.}
\label{globalGSMF}
\end{figure*}

The number of galaxies and the volume sampled by VIPERS allows us to
obtain an estimate of the GSMF with high statistical precision within
six redshifts bins in the range $0.5 \leqslant z \leqslant 1.3$.
Given the large number of galaxies observed by VIPERS, in terms of
Poisson noise it would be possible to choose even narrower bins
(e.g.~$\Delta z \simeq 0.05$ wide).  However, in that case the
measurements start being strongly affected by cosmic (sample)
variance.  A more detailed discussion is given in Sect.~\ref{Cosmic
  variance}.

We compute the GSMF within each redshift bin, using the classical
non-parametric $1/V_\mathrm{max}$ estimator \citep{Schmidt1968}.  With
this method, the density of galaxies in a given stellar mass bin is
obtained as the sum of the inverse of the volumes in which each galaxy
would be observable, multiplied by the statistical weight described in
Sect.~\ref{Spectroscopy}.  To optimise the binning in stellar mass, we
use an adaptive algorithm that extends the width of a bin until it
contains a minimum of three objects.  The errors associated with the
$1/V_\mathrm{max}$ estimates are computed assuming Poisson statistics
and include statistical weights.  The upper limits for non-detections
have been estimated following \citet{Gehrels1986}.  The values of the
$1/V_\mathrm{max}$ GSMF and associated Poisson errors are given in
Table~\ref{vmax_tab}.

It is well known that the $1/V_\mathrm{max}$ estimator is unbiased in
case of a homogeneous distribution of sources \citep{Felten1976}, but
it is affected by the presence of clustering \citep{Takeuchi2000}.  At
variance with the data sets on which the estimator was tested in the
past, VIPERS has a specific advantage, thanks to its large volume over
two independent fields. The competing effects of over- and under-dense
regions on the estimate should cancel out in such a situation.  The
impact on our analysis will also be negligible because an
inhomogeneous distribution of sources mainly affects the faint end
(i.e.~the low mass end) of the luminosity (stellar mass) function
\citep{Takeuchi2000}, while we are mainly interested in the massive
tail of the distribution.

To verify this, we compare the $1/V_\mathrm{max}$ estimates with those
of a different estimator \citep[i.e.~the stepwise maximum-likelihood
method of][]{Efstathiou1988} from another software package
\citep[ALF,][]{Ilbert2005}.  We find no significant differences in the
obtained mass functions, within the stellar mass range considered in
the present study.

Finally, in addition to the non-parametric method, we fit a
\citet{Schechter1976} function, that is,
\begin{equation}
 \Phi(\mathcal{M})\mathrm{d}\mathcal{M} =
  \Phi_\star \left( \frac{\mathcal{M}}{\mathcal{M}_\star} \right)^{\alpha} 
\exp\left( - \frac{\mathcal{M}}{\mathcal{M}_\star} \right) 
 \frac{\mathrm{d}\mathcal{M}}{\mathcal{M}_\star}  \: ,
\label{Schechter}
\end{equation}
to the $1/V_\mathrm{max}$ estimates.  The results are shown in
Fig.~\ref{globalGSMF} and in Table~\ref{sch_tab}.  Although the mass
function does not show any evidence of a rapid decline below the
completeness limit \citep[as in][]{Drory2009}, points beyond this
threshold should be considered as conservative lower limits.  These
plots clearly show the statistical power of the VIPERS sample, which
includes a significant number of the rare massive galaxies that
populate the GSMF high-mass end, thanks to its large volume.

\begin{figure}[b!]
\centering
\includegraphics[width=0.49\textwidth]{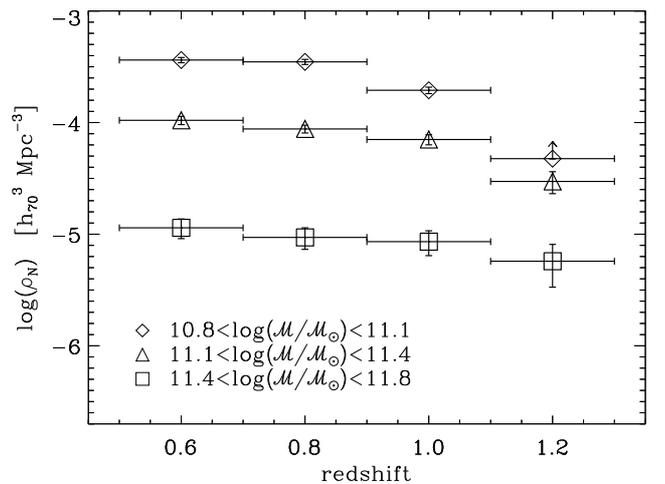}
\caption{Evolution of the galaxy number density in different bins of
  stellar mass.  The error bars of the density estimates include
  Poisson noise and cosmic variance (see Sect.~\ref{Cosmic variance}).
  At $z\simeq1.2$, for the lowest mass sample, only a lower limit can
  be estimated, indicated by the arrow. }
\label{ND}
\end{figure}

At $z < 0.6$ there is some hint of the characteristic dip of the mass
function at $\log(\mathcal{M}/\mathcal{M}_\odot)\sim 10.2$, with an
upturn below that value as observed both locally
\citep[e.g.][]{Baldry2012} and at intermediate redshifts
\citep[e.g][]{Drory2009,Pozzetti2010}.  However, this feature is
located too close to $\mathcal{M}_\mathrm{lim}$ to be assessed
effectively. We avoid using a double Schechter function in our fits
also to ease comparison with the parameters derived at higher
redshifts.  In fitting the points in the first bin ($0.5<z<0.6$), all
parameters of Eq.~\ref{Schechter} are left free, obtaining a value of
the slope $\alpha=-0.95$.  Above this redshift, however, the slope of
the low-mass end is only weakly constrained, given the relatively high
values of the completeness limit $\mathcal{M}_\mathrm{lim}$.  For this
reason, in all the other bins we fix $\alpha$ to the value $-0.95$
(see Table \ref{sch_tab}).

%
\begin{table*}
\caption{Global GSMF: $1/V_{\rm max}$ values in regular bins of stellar mass.}
\label{vmax_tab}
\setlength{\extrarowheight}{1ex}
\centering
\begin{tabular}{lcccccc}
\hline\hline
$\log\mathcal{M}\;[h_{70}^{-2}\,\mathcal{M}_\odot]$ & \multicolumn{6}{c}{$\log\Phi\;[h_{70}^{3}\, \mathrm{Mpc}^{-3}]$} \\
 & $0.5<z<0.6$ & $0.6<z<0.7$ & $0.7<z<0.8$ & $0.8<z<0.9$& $0.9<z<1.1$ & $1.1<z<1.3$ \\ [1ex]
\hline
 $ 9.50$ & $-2.47^{+0.02}_{-0.02}$ & $-2.42^{+0.01}_{-0.02}$ & $-2.62^{+0.02}_{-0.02}$ & $-2.89^{+0.03}_{-0.03}$ & $-3.20^{+0.04}_{-0.04}$ & $-4.11^{+0.11}_{-0.15}$  \\  
 $ 9.70$ & $-2.51^{+0.02}_{-0.02}$ & $-2.49^{+0.01}_{-0.01}$ & $-2.56^{+0.02}_{-0.02}$ & $-2.78^{+0.02}_{-0.02}$ & $-2.95^{+0.03}_{-0.03}$ & $-3.95^{+0.07}_{-0.08}$  \\  
 $ 9.90$ & $-2.61^{+0.02}_{-0.02}$ & $-2.56^{+0.02}_{-0.02}$ & $-2.63^{+0.02}_{-0.02}$ & $-2.79^{+0.03}_{-0.03}$ & $-2.93^{+0.03}_{-0.03}$ & $-3.69^{+0.06}_{-0.07}$  \\  
 $10.10$ & $-2.67^{+0.02}_{-0.02}$ & $-2.59^{+0.02}_{-0.02}$ & $-2.65^{+0.02}_{-0.02}$ & $-2.84^{+0.02}_{-0.02}$ & $-2.98^{+0.03}_{-0.03}$ & $-3.58^{+0.07}_{-0.08}$  \\
 $10.30$ & $-2.68^{+0.02}_{-0.02}$ & $-2.59^{+0.01}_{-0.01}$ & $-2.69^{+0.02}_{-0.02}$ & $-2.85^{+0.02}_{-0.02}$ & $-3.07^{+0.03}_{-0.03}$ & $-3.53^{+0.06}_{-0.07}$  \\ 
 $10.50$ & $-2.66^{+0.02}_{-0.02}$ & $-2.62^{+0.01}_{-0.01}$ & $-2.70^{+0.02}_{-0.02}$ & $-2.85^{+0.02}_{-0.02}$ & $-3.07^{+0.03}_{-0.03}$ & $-3.73^{+0.05}_{-0.05}$  \\
 $10.70$ & $-2.72^{+0.02}_{-0.02}$ & $-2.67^{+0.01}_{-0.01}$ & $-2.75^{+0.01}_{-0.02}$ & $-2.83^{+0.02}_{-0.02}$ & $-3.04^{+0.02}_{-0.02}$ & $-3.74^{+0.10}_{-0.13}$  \\ 
 $10.90$ & $-2.91^{+0.02}_{-0.02}$ & $-2.81^{+0.02}_{-0.02}$ & $-2.83^{+0.02}_{-0.02}$ & $-2.97^{+0.02}_{-0.02}$ & $-3.16^{+0.02}_{-0.02}$ & $-3.71^{+0.06}_{-0.07}$  \\ 
 $11.10$ & $-3.25^{+0.03}_{-0.03}$ & $-3.11^{+0.02}_{-0.02}$ & $-3.14^{+0.02}_{-0.02}$ & $-3.26^{+0.02}_{-0.03}$ & $-3.32^{+0.02}_{-0.03}$ & $-3.93^{+0.07}_{-0.09}$  \\ 
 $11.30$ & $-3.66^{+0.05}_{-0.05}$ & $-3.55^{+0.04}_{-0.04}$ & $-3.59^{+0.04}_{-0.04}$ & $-3.83^{+0.04}_{-0.05}$ & $-3.81^{+0.04}_{-0.04}$ & $-4.13^{+0.09}_{-0.12}$  \\
 $11.50$ & $-4.34^{+0.09}_{-0.12}$ & $-4.22^{+0.07}_{-0.09}$ & $-4.29^{+0.07}_{-0.09}$ & $-4.54^{+0.09}_{-0.12}$ & $-4.39^{+0.07}_{-0.08}$ & $-4.65^{+0.11}_{-0.15}$  \\ 
 $11.70$ & $-5.29^{+0.23}_{-0.53}$ & $-5.69^{+0.30}_{{\rm -inf}}$ & $-5.05^{+0.16}_{-0.26}$ & $-5.19^{+0.18}_{-0.30}$ & $-5.78^{+0.23}_{-0.54}$ & $-5.20^{+0.14}_{-0.21}$ \\ [1ex]
\hline 
\end{tabular}
\end{table*}

The results of Fig.~\ref{globalGSMF} confirm, with impressive
statistical precision, the lack of evolution since $z\simeq 1.1$ of
the massive end ($\log(\mathcal{M}/\mathcal{M}_\odot)>11$) of the
galaxy mass function seen in previous, smaller samples.  The
exponential tail of the Schechter fit is nearly constant across the
five redshift bins, down to $z\simeq 0.5$ (see Fig.~\ref{globalGSMF}).
However, we detect a significant decrease in the number density of the
most massive galaxies ($\log(\mathcal{M/M}_\odot)>11.1$) in the
redshift bin $z=1.1<z<1.3$.  At lower masses
($10.8<\log(\mathcal{M}/\mathcal{M}_\odot)<11.1$), the first signs of
evolution with respect to $z\sim 0.5$ start to be visible at redshift
$0.9$ -- $1.1$.

These trends are shown better in Fig.~\ref{ND}, where the number
density of galaxies $\rho_N$ within three mass ranges is plotted
versus redshift.  This figure explicitly shows that the most massive
galaxies are virtually already in place at $z\simeq 1$. In contrast,
galaxies with lower mass keep assembling their stars in such a way
that their number density increases by a factor $\sim 3.5$ from
$z=1.2$ down to $0.6$, consistently with the so-called
\textit{downsizing} scenario \citep{Cowie1996, Fontanot2009}.  These
new measurements confirm previous evidence, but with higher
statistical reliability (see Sect.~\ref{Comparisons to previous
  work}).

%
\begin{table}
\caption{Global GSMF: Schechter parameters ($\alpha$ fixed at $z>0.6$).}
\label{sch_tab}
\setlength{\extrarowheight}{1ex}
\centering
\begin{tabular}{lccc}
\hline\hline 
$z$ range & $\alpha$ & $\log\mathcal{M}_\star$       & $\Phi_\star$ \\
          &          & $[h_{70}^{-2}\,\mathcal{M}_\odot]$ & $[10^{-3}\,h_{70}^3\,\mathrm{Mpc}^{-3}]$ \\ [1ex]
\hline
$0.5-0.6$ &  $-0.95^{+0.03}_{-0.02}$ & $10.87^{+0.02}_{-0.02}$ & $1.42^{+0.06}_{-0.07}$ \\
$0.6-0.7$ &  $-0.95$ & $10.91^{+0.02}_{-0.01}$ & $1.58^{+0.05}_{-0.05}$ \\
$0.7-0.8$ &  $-0.95$ & $10.91^{+0.01}_{-0.02}$ & $1.38^{+0.06}_{-0.04}$ \\
$0.8-0.9$ &  $-0.95$ & $10.85^{+0.02}_{-0.02}$ & $1.29^{+0.09}_{-0.09}$ \\
$9.0-1.1$ &  $-0.95$ & $10.91^{+0.02}_{-0.01}$ & $0.82^{+0.05}_{-0.06}$ \\
$1.1-1.3$ &  $-0.95$ & $11.03^{+0.11}_{-0.08}$ & $0.20^{+0.05}_{-0.06}$ \\ [1ex]
\hline
\end{tabular}
\end{table}
%

\subsection{Cosmic variance in the VIPERS survey}
\label{Cosmic variance}

\begin{figure}
\centering
\includegraphics[width=0.48\textwidth]{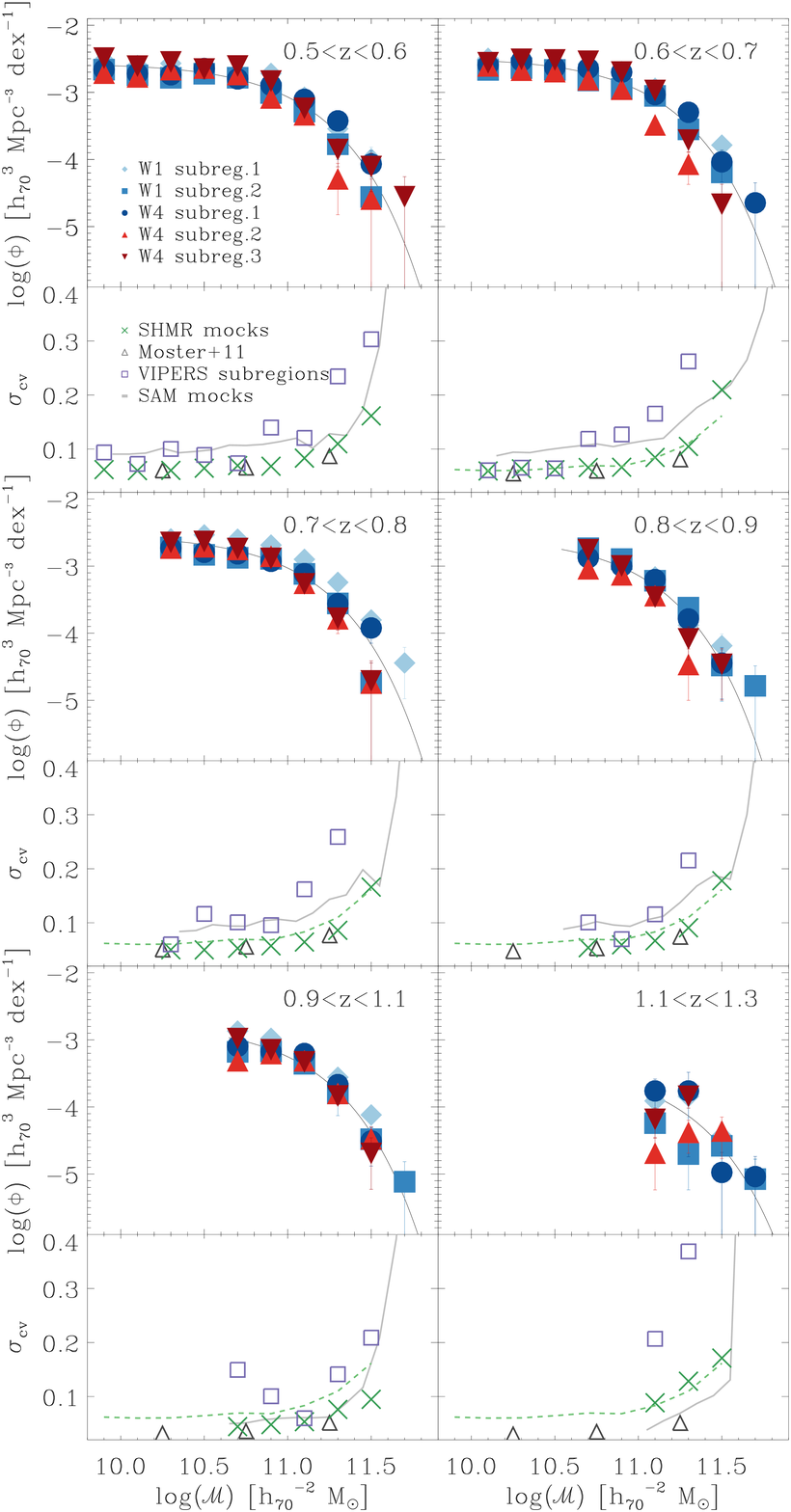}
\caption{Estimates of the contribution of sample (cosmic) variance to
  the statistical uncertainty of the GSMF measurements. For each
  redshift bin, the upper panels show the GSMF $1/V_\mathrm{max}$
  measurements obtained from five VIPERS subregions of
  $2\,\mathrm{deg}^2$, located respectively in the W1 field (three
  regions, blue diamonds, circles, and squares) and in the W4 field
  (two regions, red triangles, and downward triangles).  The Schechter
  fit to the global GSMF of Fig.~\ref{globalGSMF} is shown as
  reference (black solid line).  The lower panels show the standard
  deviations estimated in each redshift bin from these five
  measurements (purple squares, Eq.~\ref{sigma_obs}), together with
  the estimates of $\sigma_\mathrm{cv}$ obtained from $57$ SHMR mocks
  by means of Eq.~\ref{sigmacv_th} (green crosses). To highlight how
  the effect of cosmic variance decreases at higher $z$, we report
  $\sigma_\mathrm{cv,SHMR}$ of the first redshift bin in the other
  panels (green dashed lines).  In addition, the sample variance
  measured in $50$ SAM mocks (grey solid line) and the estimates
  provided by \citet{Moster2011} method (black triangles) are shown as
  reference.}
\label{cvplot}
\end{figure}

When dealing with statistical studies using number counts, a severe
complication is introduced by the field-to-field fluctuations in the
source density, due to the clustered nature of the galaxy distribution
and the existence of fluctuations on scales comparable to the survey
volume.  This sampling or `cosmic' variance represents a further term
of uncertainty to be added to the Poisson shot noise.  It can be
expressed by removing $\sigma_\mathrm{Poiss}^2 \equiv 1/\langle N
\rangle$ from the total relative error:
\begin{equation}
\sigma_\mathrm{cv}^2 = \frac{\langle N^2 \rangle - \langle N \rangle^2}{\langle N \rangle^2} - \frac{1}{ \langle N \rangle}  ,
\label{sigmacv_th}
\end{equation}
where $\langle N \rangle$ and $ \langle N^2 \rangle $ are the mean and
the variance of galaxy number counts \citep{Somerville2004}.

Extragalactic pencil-beam surveys, even the deepest ones, are
particularly limited by cosmic variance, given the small volume
covered per redshift interval.  At $z \sim 0.8$, galaxy density
fluctuations are found to be still relevant up to a scale of $\sim
140\,{\rm Mpc}\,h_{70}^{-1}$ \citep{Scrimgeour2012}, which roughly
corresponds to $5\,\mathrm{deg}$.

This is the result of intrinsic clustering in the matter, as predicted
by the power spectrum shape and amplitude at that epoch, amplified by
the bias factor of the class of galaxies analysed, which at high
redshift can be very large for some classes.  Also the
last-generation, largest deep surveys are significantly affected by
this issue.  For example, the COSMOS field, despite its
$2\,\mathrm{deg}^2$ area, turned out to be significantly overly dense  
between $z=0.8$ and $z=1$ \citep{Kovac2010a}.

The gain obtained by enlarging the area of a single field beyond a
certain coverage becomes less prominent, owing to the existing
large-scale correlations \citep[see][Fig.~1]{Newman&Davis2002}:
$\sigma_\mathrm{cv}$ decreases mildly as a function of volume, with an
approximate dependence $\sigma_\mathrm{cv}\propto V^{-0.3}$
\citep[][Fig.~2]{Somerville2004}, compared to $\sigma_\mathrm{Poiss}
\propto V^{-0.5}$. \citet{Trenti&Stiavelli2008} found similar results
by characterizing Lyman break galaxies surveys: at high values of
$\langle N \rangle$, the Poisson noise rapidly drops and cosmic
variance remains the dominant source of uncertainty.  A more effective
way to abate cosmic variance is to observe separated regions of sky.
Since counts in these regions, if they are sufficiently distant, are
uncorrelated, their variances sum up in quadrature (i.e.,
$\sigma_\mathrm{cv}$ decreases as the square root of the number of
fields, \citealp{Moster2011}).  Multiple independent fields can then
result in a smaller uncertainty than for a single field, even if the
latter has a larger effective area \citep{Trenti&Stiavelli2008}.  The
current VIPERS PDR-1 sample is not only characterised by a
significantly large area, compared to previous similar surveys at
these redshifts, but it is also split into two independent and
well-separated fields of $\sim 7.5\,\mathrm{deg}^2$ each. We therefore
expect that the impact of cosmic variance should be limited.

To quantify this effect directly, we follow two approaches.
The first one, based on the observations themselves, provides an upper
limit of the VIPERS $\sigma_\mathrm{cv}$.  We select five rectangular
subregions of about $2\,\mathrm{deg}^2$ within the survey and estimate
the mass function $\Phi_i$ in each of them, using the
$1/V_\mathrm{max}$ method described above.  We choose non-contiguous
regions (separated by $\sim1\,\mathrm{deg}$) to minimise the
covariance between subsamples located within the same field (W1 or
W4).  Within mass bins $\mathcal{M}_j\pm\Delta\mathcal{M}/2$ we derive
the total random uncertainty
\begin{equation}
\sigma_\mathrm{tot, obs}(\mathcal{M}_j) =\frac{1}{n} \sum_{i=1}^n \sqrt{ \left[ \Phi_i(\mathcal{M}_j) - \Phi_\mathrm{tot}(\mathcal{M}_j) \right]^2 }  \,\,\,\, ,
\label{sigma_obs}
\end{equation}  
where $\Phi_\mathrm{tot}$ is the global GSMF of VIPERS (at that
redshift) and $\Phi_i(\mathcal{M}_j)$ the number density of galaxies
measured in the $j$-th mass bin for each of the $n=5$ subregions.
This result should be regarded as an upper limit of the VIPERS cosmic
variance, given that the subsamples have a smaller volume than the
whole survey, and Eq.~\ref{sigma_obs} also includes the variance due
to Poisson noise.  Conversely, residual correlation among the
subfields within each of the VIPERS fields (produced by structures on
scales $\gtrsim 1\,\mathrm{deg}$ crossing over two or more subregions)
would slightly reduce $\sigma_\mathrm{tot, obs}$.  More in general,
the small number of fields used to perform this test makes the
computation of Eq.~\ref{sigma_obs} statistically uncertain: for these
reasons the estimates of the standard deviation obtained from the
field-to-field fluctuations among the five subsamples
($\sigma_\mathrm{tot, obs}$, squares in Fig.~\ref{cvplot}) show rather
irregular behaviour.

The second approach is based on the use of simulated mock surveys.
First, we use a set of $57$ mock samples (26 and 31 in W1 and W4,
respectively), built using specific recipes for the stellar-to-halo
mass relation.  They are based on the MultiDark dark matter simulation
\citep{Prada2012} and have been constructed to reproduce the detailed
geometry and selection function of the VIPERS survey up to $z=1.2$.
\citep[see][for details]{delaTorre2013}. The dark matter haloes
identified in the simulation, as well as artificial sub-haloes drawn
from the \citet{Giocoli2010} subhalo mass function, have been
associated with galaxies using the stellar-to-halo mass relations of
\citet{Moster2013}.  The latter are calibrated on previous stellar
mass function measurements in the redshift range $0 < z < 4$.  We call
these `SHMR mocks'.  We apply Eq.~\ref{sigmacv_th} to estimate the
amount of cosmic variance independently among the 26 W1 and 31 W4
mocks.  The global estimate of cosmic variance ($\sigma_\mathrm{cv,
  SHMR}$) on the scales of the VIPERS survey is obtained by combining
the results from the two fields \citep[see][Eq.~7]{Moster2011}.  As
expected, we find that $\sigma_\mathrm{cv,SHMR}$ decreases with
redshift, since we are probing larger and larger volumes, and
increases with stellar mass owing to the higher bias factor (and thus
higher clustering) of massive galaxies \citep{Somerville2004}.  Both
trends are clearly visible in Fig.~\ref{cvplot}, where measurements of
$\sigma_\mathrm{cv, SHMR}$ are presented for different bins of
redshift and stellar mass. These values are included in the error bars
of Fig.~\ref{globalGSMF} to account for the cosmic variance
uncertainty. We notice that in the highest redshift bin
$\sigma_\mathrm{cv, SHMR}$ represents a conservative estimate, given
the different redshift range in SHMR mocks ($1.1<z<1.2$) and
observations ($1.1<z<1.3$).

In Fig.~\ref{cvplot} we also show, as a reference, the estimates
provided by the public code {\tt getcv} \citep{Moster2011} for the
same area of the SHMR mocks. These results, limited at
$\log(\mathcal{M}/\mathcal{M}_\odot)\leqslant11.5$, are in good
agreement with $\sigma_\mathrm{cv,SHMR}$, with the exception of the
highest redshift bin, mainly because of the $z=1.2$ cut of SHMR mocks.
However, we prefer to use $\sigma_\mathrm{cv, SHMR}$ to quantify the
cosmic variance uncertainty in that $z$-bin, although it should be
regarded as an upper limit, since the outcomes of \citet{Moster2011}
code do not reach the high-mass tail of the GSMF, and are also more
uncertain because the galaxy bias function used in this method is less
constrained at such redshifts.

Besides these SHMR mocks, we also used another set of $50$ VIPERS-like
light cones built from the Millennium simulation \citep{Springel2005},
in which dark-matter haloes are populated with galaxies through the
semi-analytical model (SAM) of \citet{DeLucia&Blaizot2007}.  Galaxy
properties were determined by connecting the astrophysical processes
with the mass accretion history of the simulated dark matter haloes.
Each mock sample covers $7 \times 1 \,\mathrm{deg}^2$, with a
magnitude cut in the $i$ band equal to that of the observed sample.
Although the geometry of these mocks (and therefore their volume)
differs slightly from the design of the real survey, they provide an
independent test, with a completely different prescription for galaxy
formation.  With respect to the SHMR mocks, SAM mocks in
Fig.~\ref{cvplot} show a trend similar to that of $\sigma_\mathrm{cv,
  SHMR}$, although with some fluctuations e.g.~between $z=0.7$ and
0.8.  The values are systematically higher mainly because the SAM
mocks do not reproduce two independent fields.  Further differences
with respect to the other estimates may be due to the different
recipes in the simulations.

\subsection{Other sources of uncertainty}
\label{Other uncertainties}

In describing our procedure to derive stellar masses by means of the
SED fitting technique (Sect.~\ref{Stellar masses}), we emphasised the
number of involved parameters and their possible influence on the
estimates.  The assumptions that have the strongest impact on the
results are the choices of the stellar population synthesis model,
IMF, SFH, metallicity, and dust extinction law.  A thorough discussion
about each one of the mentioned ingredients is beyond the goals of
this paper, but the reader is referred to \citet{Conroy2013},
\citet{Mitchell2013}, and \citet{Marchesini2009} for a comprehensive
review of the systematic effects induced by the choice of the input
parameters.

Here we briefly test the impact on the GSMF of choosing different
values of $Z$ (whether including subsolar metallicities or not), the
extinction laws (SB and SMC, or SB alone), and the addition of
secondary bursts to the smooth SFHs (i.e.~complex SFHs instead of
exponentially declining $\tau$-models).  We do not modify the other
two main ingredients in our procedure, i.e.~the universal IMF that we
assumed \citep{Chabrier2003} and the stellar population synthesis
model (BC03).
 
To perform this test we use stellar mass estimates obtained by
assuming five different sets of SED fitting templates, four of them
differing in metallicity and extinction law: $Z_\odot$ only and SB;
two metallicities ($Z_\odot$ and $0.2Z_\odot$) and SB; solar
metallicity and two extinction laws (SB and SMC); two metallicities
($Z_\odot$ and $0.2 Z_\odot$) and two extinction laws (SB and SMC).
The fifth SED fitting estimate has been derived with the
\textit{MAGPHYS} code (see Sect.~\ref{Stellar masses}), assuming the
following parameters: complex SFHs, extinction model derived from
\citet{Charlot&Fall2000}, and a wider range of metallicity (including
super-solar ones).  We limit these tests to the data in the VIPERS W1
field, i.e. about half of the total sample, given the better overall
photometric coverage in this area and the large computational time
involved.

\begin{figure}
\centering
\includegraphics[width=0.49\textwidth]{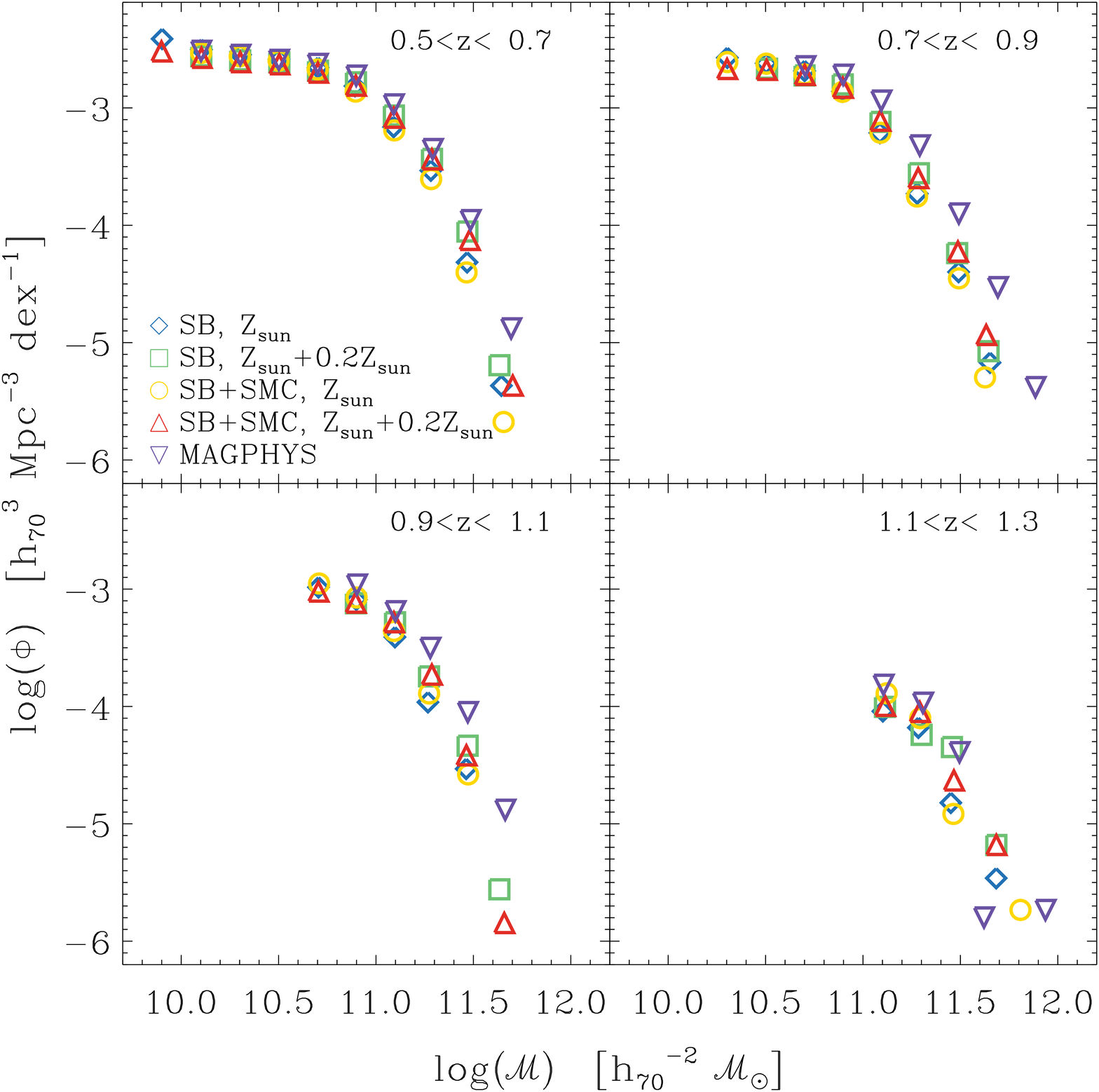}
\caption{Dependence of the mass function on the details of the stellar
  mass estimates, considering five different cases.  Specifically, the
  points correspond to different choices of the stellar population
  synthesis code, metallicity ($Z$), extinction law (SB+SMC or SB
  alone), or the addition of secondary bursts to the smooth
  star-formation histories.  Four cases correspond to SED fitting
  using \textit{Hyperzmass}, for which the values of the adopted
  parameters are given in the bottom-left of the first panel.  For
  details about the parameters adopted for \textit{MAGPHYS} (downward
  triangle), we refer to Sect.~\ref{Stellar masses}.}
\label{SEDcf} 
\end{figure}

The mass functions resulting from these five different SED-modelling
assumptions are shown in Fig.~\ref{SEDcf}.  As expected (see
discussion in Sect.~\ref{Stellar masses} and Fig.~\ref{2bursts}), the
\textit{MAGPHYS} mass function corresponds to the highest estimated
values of galaxy density at high stellar masses (at least up to
$z\simeq 1.1$).  This trend is expected, because the four other
estimates, obtained by assuming smooth SFHs templates, are insensitive
to an underlying old stellar population that is outshone by a recent
burst of star formation (\citealp{Fontana2004,Pozzetti2010}, but see
\citealp{Moustakas2013} for an opposite result).  As a consequence,
when using complex SFHs templates one can produce stellar mass
estimates that are higher than those obtained with smooth SFHs for a
low percentage of objects, an effect that is more evident in the
high-mass tail.
 
The other estimates, produced by \textit{Hyperzmass}, are in quite
good agreement with each other. The mass functions are slightly higher
(on average by about $0.1$\,dex) when obtained through SED fitting
procedures that can choose between two values of metallicity. In fact,
in this case, red galaxies can be fit with $0.2\,Z_\odot$ and older
ages, consequently resulting in higher stellar mass values. The effect
of the extinction law is instead marginal.

\section{Comparison to previous work}
\label{Comparisons to previous work}

In this section we compare the VIPERS GSMF with other mass functions
derived from different galaxy surveys (Sect.~\ref{Comparison with
  observations}) and various semi-analytical models
(Sect.~\ref{Comparison with models}).

\begin{figure*}
  \centering
  \includegraphics[width=0.9\textwidth]{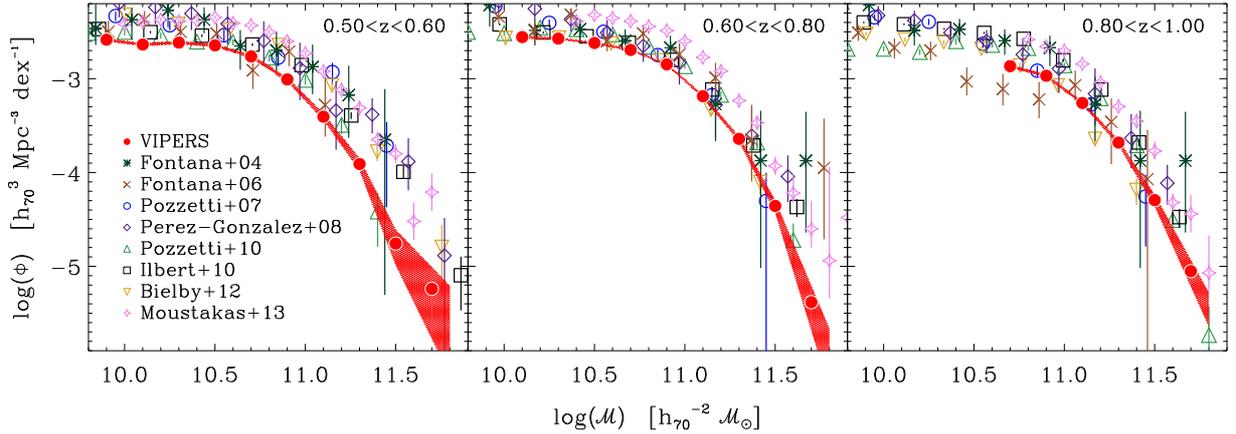}
  \caption{The VIPERS galaxy stellar mass functions from $z=0.5$ to
    $1$ (filled red circles, with a red shaded area accounting for the
    Poisson uncertainty). The $1/V_\mathrm{max}$ determinations of
    previous surveys are also shown by different symbols, along with
    their respective Poisson error bars. In the left-hand panel,
    whereas the VIPERS range is $0.5<z<0.6$, the other GSMFs are
    computed between $z=0.4$ and $0.6$, with the exception of
    \citet[][]{Moustakas2013} for which is $0.5<z<0.65$, $0.65<z<0.8$,
    $0.8<z<1.0$; notice the very small error bars of the VIPERS data,
    despite the narrower redshift range.  In the other two panels, the
    bins of VIPERS are the same as the other surveys; also at these
    higher redshifts the error bars of the VIPERS GSMF are
    small compared to them.}
  \label{globalGSMFcomp}
\end{figure*}

\subsection{Comparison with other observational estimates}
\label{Comparison with observations}

We compare here our estimate of the GSMF with results from other
galaxy surveys.  We correct GSMFs (if necessary) to be in the same
cosmological model with $\Omega_m=0.3$, $\Omega_\mathbf{\Lambda}=0.7$,
$h_{70}=1$, and \citet{Chabrier2003} IMF.  We also modify our binning
in redshift to be similar to other work.

We chose eight surveys that adopt comparable $z$-bins, half of them
based on photometric redshifts
\citep{Fontana2006,PerezGonzalez2008,Ilbert2010,Bielby2012} and half
on spectroscopic redshifts
\citep{Fontana2004,Pozzetti2007,Pozzetti2010,Moustakas2013}.  The
spectroscopic redshift sample used by \citet{Moustakas2013} is
obtained through a pioneering technique based on a low dispersion
prism and slitmasks \citep[][]{Coil2011}, which results in a precision
of $\sigma_z\simeq 0.007 (1+z)$ \citep[for their high quality sample
$Q \geqslant 3$, see][]{Cool2013}, i.e. comparable to the precision of
the best photometric redshifts \citep[][who obtain $\sigma_z \simeq
0.008 (1+z)$ and a very low percentage of outliers]{Ilbert2013}.

The redshift ranges of the GSMFs shown in Fig.~\ref{globalGSMFcomp}
are $0.4<z<0.6$, $0.6<z<0.8$, $0.8<z<1.0$, with the exception of
PRIMUS \citep{Moustakas2013}, which is at $0.5<z<0.65$, $0.65<z<0.8$,
$0.8<z<1.0$, and the first bin of VIPERS (i.e., $0.5<z<0.6$).  In the
case of \citet{Bielby2012}, who provide the GSMFs in four CFHTLS-Deep
quadrants, we consider the results in the D3 field
($1\,440\,\mathrm{arcmin}^2$), which is located in a region of sky
uncorrelated with the other surveys we selected.  For the VIPERS GSMFs
we plot error bars accounting only for $\sigma_\mathrm{Poiss}$,
i.e.~without adding the uncertainty due to sample variance, in order
to be consistent with most of the literature data, for which only
Poisson errors are available.
\footnote{Nonetheless, through the recipe of \citet{Moster2011} we can
  obtain, for each survey, an approximate estimate of the uncertainty
  due to cosmic variance to a first approximation, and have a rough
  idea of how much the error bars would increase in
  Fig.~\ref{globalGSMFcomp} when accounting for it.  For
  \citet{Pozzetti2007}, \citet{PerezGonzalez2008}, and
  \citet{Bielby2012}, within the redshift ranges considered in
  Fig.~\ref{globalGSMFcomp}, with only a small evolution with
  redshift, the GSMF uncertainty related to cosmic variance is
  approximatively the same: $\sim15\%$ between
  $\log\mathcal{M/M_\odot}=10.0$ and 10.5, $\sim23\%$ between
  $\log\mathcal{M/M_\odot}=11.0$ and 11.5. (It should be noticed that
  data used by \citeauthor{PerezGonzalez2008} cover an area of
  $273\,\mathrm{arcmin}^2$, but split in three fields.)  For
  \citet{Ilbert2010} and \citet{Pozzetti2010},
  $\sigma_\mathrm{cv}\simeq10\%$ when
  $10.0<\log\mathcal{M/M_\odot}<10.5$ and
  $\sigma_\mathrm{cv}\simeq17\%$ when
  $11.0<\log\mathcal{M/M_\odot}<11.5$. In the same bins of stellar
  mass, for \citet{Fontana2004} $\sigma_\mathrm{cv}$ is $20\%$ and
  $30\%$, respectively, while $\sigma_\mathrm{cv}\simeq30\%$ and 45\%
  in \citet{Fontana2006}. The estimates provided by
  \citet{Moustakas2013} in their paper are generally below 10\%,
  except at $\log\mathcal{M/M_\odot}>11.6$ where the uncertainty rises
  by a factor of $2-4$.} 

Our results lie on the lower boundary of the range
covered by other GSMFs, and are in reasonably good agreement with most
of them. At $0.8<z<1.0$, the difference with \citet[][COSMOS survey over
$2\,\mathrm{deg}^2$]{Ilbert2010} and \citet[][zCOSMOS,
$1.4\,\mathrm{deg}^2$]{Pozzetti2010}  is
noteworthy: the likely reason is the presence of a large structure
detected in the COSMOS/zCOSMOS field \citep{Kovac2010a}, demonstrating
the importance of the cosmic variance in this kind of comparison.

Some discrepancy (nearly by a factor of two) is also evident with the
estimates by \citet{Moustakas2013}.  The explanation could be partly
related to the statistical weighing, in particular for the faintest
objects, because the lower the sampling rate estimates, the greater
the uncertainty in such a correction.  At magnitudes $i \simeq 22.5$,
the SSR of PRIMUS is approximately $45$\%, dropping below $20$\% at
the limit of the survey \citep{Cool2013}.  Instead, in VIPERS the SSR
is $\sim 75$\% down to our magnitude limit $i=22.5$ and to $z\simeq
1$, making the statistical weight corrections smaller and more robust.
In addition to this, it should be noticed that although several
overdensities have been observed in PRIMUS, cosmic variance seems
unable to fully justify the difference between the GSMFs of the two
surveys: the number of independent fields (PRIMUS consists of five
fields with a total of $5.5\,\mathrm{deg}^2$) should reduce this
problem, at least to some degree.  The 
disagreement could also be partially ascribed to the different ways
stellar masses are estimated: \citeauthor{Moustakas2013} derived their
reference SEDs according to the SSP model of \citet{Conroy&Gunn2010},
which results in stellar mass estimates systematically higher than
those obtained by assuming BC03 \citep[see][Fig.~19]{Moustakas2013}.

Regarding the choices of SEDs, it is worth noticing that
\citet{PerezGonzalez2008} also used a template library different from
ours, which they derived from the PEGASE stellar population synthesis
model \citep{Fioc&RoccaVolmerange1997}, bounding the parameter space
by means of a training set of $\sim 2000$ galaxies with spectroscopic
$z$ and wide photometric baseline.  The other surveys quoted in
Fig.~\ref{globalGSMFcomp}
\citep{Fontana2004,Fontana2006,Pozzetti2007,Pozzetti2010,Ilbert2010,Bielby2012}
adopt BC03.

VIPERS data provide tight constraints on the high-mass end of the
GSMF.  Previous surveys, such as K20, MUSIC, and VVDS-Deep
\citep[i.e.][]{Fontana2004,Fontana2006,Pozzetti2007}, were unable to
probe this portion of the GSMF
($\log(\mathcal{M}/\mathcal{M}_\odot)\gtrsim 11.5$) because of their
relatively small area (about $52$, $150$, and
$1\,750\,\mathrm{arcmin}^2$ respectively). Instead, GSMFs derived from
photometric redshift surveys are characterised by a Poisson noise that
is in general comparable to the level in VIPERS
\citep{PerezGonzalez2008,Ilbert2010}, but they can be affected by
failures on photometric redshift estimates: even a small fraction of
catastrophic redshift measurements can be relevant at high masses
\citep{Marchesini2009,Marchesini2010}.  Moreover, the sky area
generally covered by high-$z$ photometric surveys is not large enough
for cosmic variance to be negligible.

We postpone a detailed analysis of the evolution of the GSMF down to
the local Universe to future work: differences in the details of the
available estimates from 2dFGRS, SDSS, and GAMA
\citep[see][]{Cole2001,Bell2003,Panter2004,Baldry2008,Li&White2009,Baldry2012}
prevent a robust comparison with our data.  Only computing stellar
masses and mass functions in a self-consistent way can provide
constraints on the evolution of the GSMF down to $z=0$
\citep[e.g.][]{Moustakas2013}.

\subsection{Testing models}
\label{Comparison with models}
%

\begin{figure}
\centering
\includegraphics[width=0.49\textwidth]{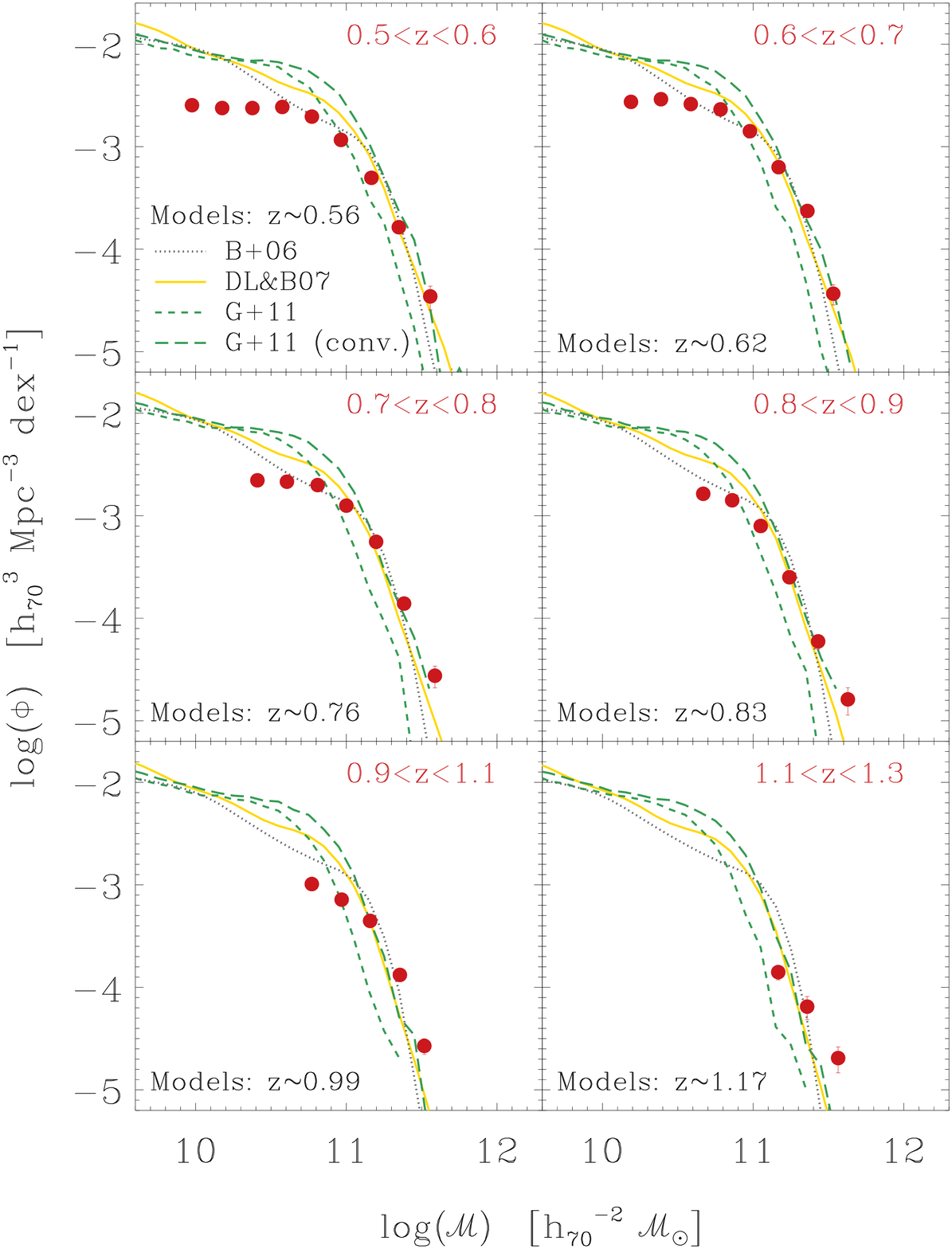}
\caption{Comparison of the VIPERS mass function (red points, as in
  Fig.~\ref{globalGSMF}) with the semi-analytical models of
  \citet{Bower2006}, \citet{DeLucia&Blaizot2007}, and \citet{Guo2011}
  (grey dotted, yellow solid, green short-dashed lines), whose GSMFs
  have been derived directly from the tables available in the
  Millennium database \citep{Lemson2006}. The \citet{Guo2011} stellar
  masses have also been convolved with a Gaussian of dispersion
  $0.15$\,dex, to reproduce observational uncertainty on stellar mass
  determinations; the resulting GSMFs are represented with green
  long-dashed lines.}
\label{model1} 
\end{figure}

Besides the comparison with other surveys, it is important to check
the agreement of our results with simulations. In this paper we limit
ourselves to a preliminary analysis. Nevertheless, this first test
provides intriguing results.

The four semi-analytical models (SAMs) we consider here rely on the
halo-merger trees of the Millennium Simulation
\citep[MS][]{Springel2005} and the Millennium-II Simulation
\citep[MSII][]{BoylanKolchin2009}; namely, three of them
\citep{Bower2006,DeLucia&Blaizot2007,Mutch2013} use the MS (comoving
box size $L=714\,\mathrm{Mpc}\,h_{70}^{-1}$, particle mass $=1.23
\times 10^9\,\mathcal{M}_\odot\,h_{70}^{-1}$), while the last one
\citep{Guo2011} is based on both MSI and MSII
($L=143\,\mathrm{Mpc}\,h_{70}^{-1}$, particle mass $=9.83 \times
10^6\,\mathcal{M}_\odot\,h_{70}^{-1}$).  The tight constraints posed
by VIPERS can be very useful when studying whether these models
adequately reproduce the real universe.

In Fig.~\ref{model1}, we show the mass functions derived from the
models of \citet{Bower2006}, \citet{DeLucia&Blaizot2007}, and
\citet{Guo2011}, together with the VIPERS results.  All the model
GSMFs are computed from snapshots at the same redshifts. The narrow
redshift binning we can set in VIPERS ($\Delta z=0.1$) allows us to
compare simulated galaxies to observed ones at cosmic times that are
very close to the snapshot considered.  In the case of
\citeauthor{DeLucia&Blaizot2007} model, we also derived the stellar
mass functions from the VIPERS-like light cones introduced in
Sect.~\ref{Cosmic variance}, but we do not show them in
Fig.~\ref{model1} since they lead to results that are
indistinguishable from those obtained from snapshots. For all three
SAMs, we find that the low-mass end of the GSMF is over-estimated.
Such a discrepancy, already observed in other work
\citep{Somerville2008,Cirasuolo2010}, is mainly due to an
over-predicted fraction of passive galaxies on those mass scales.
This can be caused by an under-efficient supernova feedback and/or
some issue as to how the star formation efficiency is parametrised at
high redshifts \citep{Fontanot2009,Guo2011}. Rescaling the simulations
to an up-to-date value of $\sigma_8$ (in MS it is equal to $0.9$),
with the consequence of reducing the small-scale clustering of
dark-matter haloes, alleviates the tension only in part
\citep{Wang2008,Guo2013}.

At a first glance, \citet{DeLucia&Blaizot2007} and \citet{Bower2006}
seem to agree with the observed GSMFs at
$\log(\mathcal{M/M_\odot})\gtrsim11.0$, while the \citet{Guo2011} mass
function lies systematically below by $\simeq 0.4$\,dex.  However, it
should be emphasised that in Fig.~\ref{model1} we plotted the GSMFs
from SAMs without taking the observational uncertainties on stellar
mass into account.  We verified that adding this kind of error would
increase the density of massive objects in the exponential tail of the
mass function, and therefore the \citet{DeLucia&Blaizot2007} and
\citet{Bower2006} results should be considered at variance with
observations also at $\log(\mathcal{M}/\mathcal{M}_\odot) \geqslant
11$.

The effect of introducing observational uncertainties is shown in
Fig.~\ref{model1} only for the \citet{Guo2011} model, which foresees a
lower density of objects in the massive end with respect to the other
two models.  We recomputed the \citeauthor{Guo2011} GSMFs after
convolving stellar masses with a Gaussian of dispersion $0.15$\,dex.
The predictions of \citet{Guo2011} are then in fair agreement with
VIPERS.  With respect to \citet{DeLucia&Blaizot2007}, the main
distinguishing features of \citet{Guo2011} model are the high
efficiency of supernova feedback and a lower rate of gas recycling at
low mass.  The transition from central to satellite status in the
\citeauthor{Guo2011} prescription also differs, resulting in a larger
number of satellite galaxies than in \citeauthor{DeLucia&Blaizot2007}
model.

\begin{figure}
\centering
  \includegraphics[width=0.49\textwidth]{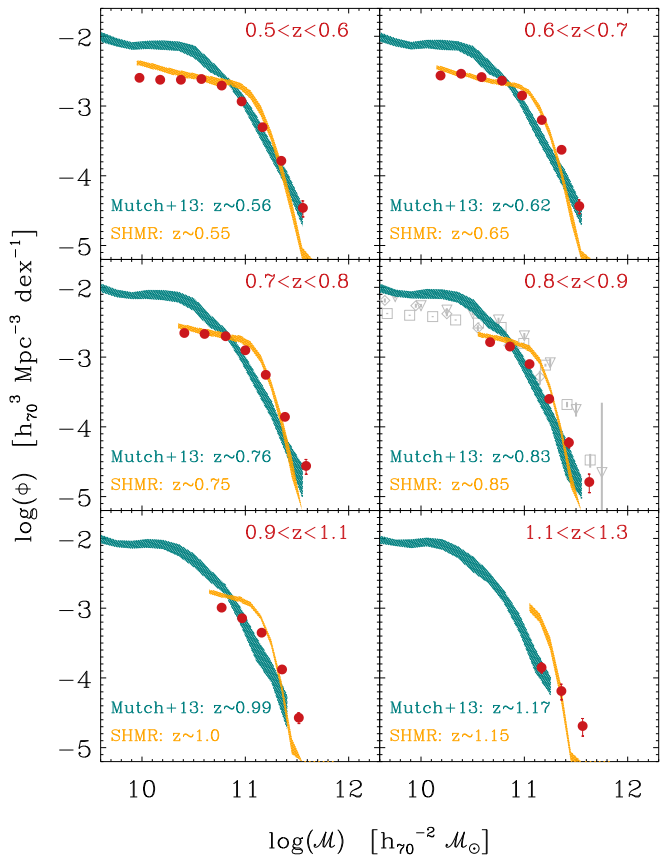}
  \caption{Comparison of the VIPERS mass function (red points) with
    the semi-analytical model of \citet{Mutch2013} (green shaded area
    at $95$\% confidence limits). In several redshift bins
    \citeauthor{Mutch2013} GSMF does not reach masses as high as
    VIPERS because the volume of the simulation (with a comoving box
    size $L=89.3\,\mathrm{Mpc}\,h_{70}^{-1}$) is smaller.  In the
    right-hand middle panel ($0.8<z<0.9$), a grey dashed
    line represents the mass function that \citeauthor{Mutch2013} obtain by
    combining observational data from three different surveys
    \citep[][grey triangles,diamonds, and squares,
    respectively]{Pozzetti2007,Drory2009,Ilbert2010}.  In addition,
    the yellow shaded regions represent the dispersion of the mass
    functions derived from the $57$ SHMR mocks (see Sect.~\ref{Cosmic
      variance}), in the same redshift bins as the VIPERS ones.}
  \label{modelcf}
\end{figure}

It should be emphasised that only \citet{Guo2011} choose most of the
parameters in order to fit the observed local mass function, whereas
\citet{Bower2006} and \citet{DeLucia&Blaizot2007} use the local
luminosity function to adjust their recipes.  In recent studies, the
parameters of these models have been tuned [again] by means of a
different approach, based on Bayesian inference
\citep{Henriques2009,Bower2010}.  From this perspective, a particular
kind of calibration has been proposed by \citet{Mutch2013}, who modify
the input parameters in the SAM of \citet{Croton2006} to match
observations at $z=0$ and $z \simeq 0.8$ simultaneously.

The results obtained by \citet{Mutch2013} are compared to the VIPERS
mass functions in Fig.~\ref{modelcf}.  The plot shows reasonable
agreement beyond $\mathcal{M} \simeq 10^{11}\,\mathcal{M}_\odot$, not
only at the redshift of calibration ($z \simeq 0.83$) but also in the
other bins.  The authors do not convolve their mass functions with a
Gaussian uncertainty on stellar masses, because at least part of the
uncertainties this procedure accounts for should already be included
in the observational constraints they use.  The \citet{Mutch2013}
model is calibrated at $z=0.83$ by using the results of
\citet{Pozzetti2007}, \citet{Drory2009}, and \citet{Ilbert2010}.
Among these three GSMFs, only \citet{Pozzetti2007} is based on
spectroscopic data (VVDS-Deep), which are unfortunately quite limited
at high masses.  The other two estimates \citep{Drory2009,Ilbert2010}
are derived from the COSMOS survey, which contains a significant
over-density at $z\simeq 0.8$.  The strategy adopted by
\citeauthor{Mutch2013} to combine such information may lead to
overconfidence in the adopted constraints, especially in the highest
mass range, where observations are most difficult.  To reconcile SAM
and observations at $\log(\mathcal{M}/\mathcal{M}_\odot) > 10.8$,
\citet{Mutch2013} have assumed a star formation efficiency much higher
than the one imposed by \citet{Croton2006}, and consequently they were
forced to parametrise supernova feedback efficiency with a range of
values that is not completely supported by observations
\citep{Rupke2002,Martin2006}.  Intriguingly, we note that the authors
would significantly relieve these tensions if they were to add VIPERS
data to their analysis.

From a different perspective, the SHMR mocks we introduced in
Sect.~\ref{Cosmic variance} are also calibrated at multiple redshifts.
We decided to test their reliability by deriving their GSMFs
(Fig.~\ref{modelcf}). The agreement is remarkable: VIPERS data confirm
the validity of the stellar-to-halo mass relation of
\citet{Moster2013} that was used to construct these mocks.  This
relation connects galaxies with their hosting dark matter halo by
means of a redshift-dependent parametrisation that has been calibrated
through the GSMFs of \citet{PerezGonzalez2008} and \citet{Santini2012}
up to $z=4$.  Because of the lack of tight constraints used by
\citeauthor{Moster2013} for the most massive galaxies (the data from
\citealp{PerezGonzalez2008} have lower statistics than ours), the SHMR
mass functions diverge at high mass from our estimates.

\section{Evolution of the mass function of the red and blue galaxy populations}
\label{Analysis by galaxy type}

\begin{figure*}
\centering
  \includegraphics[width=0.99\textwidth]{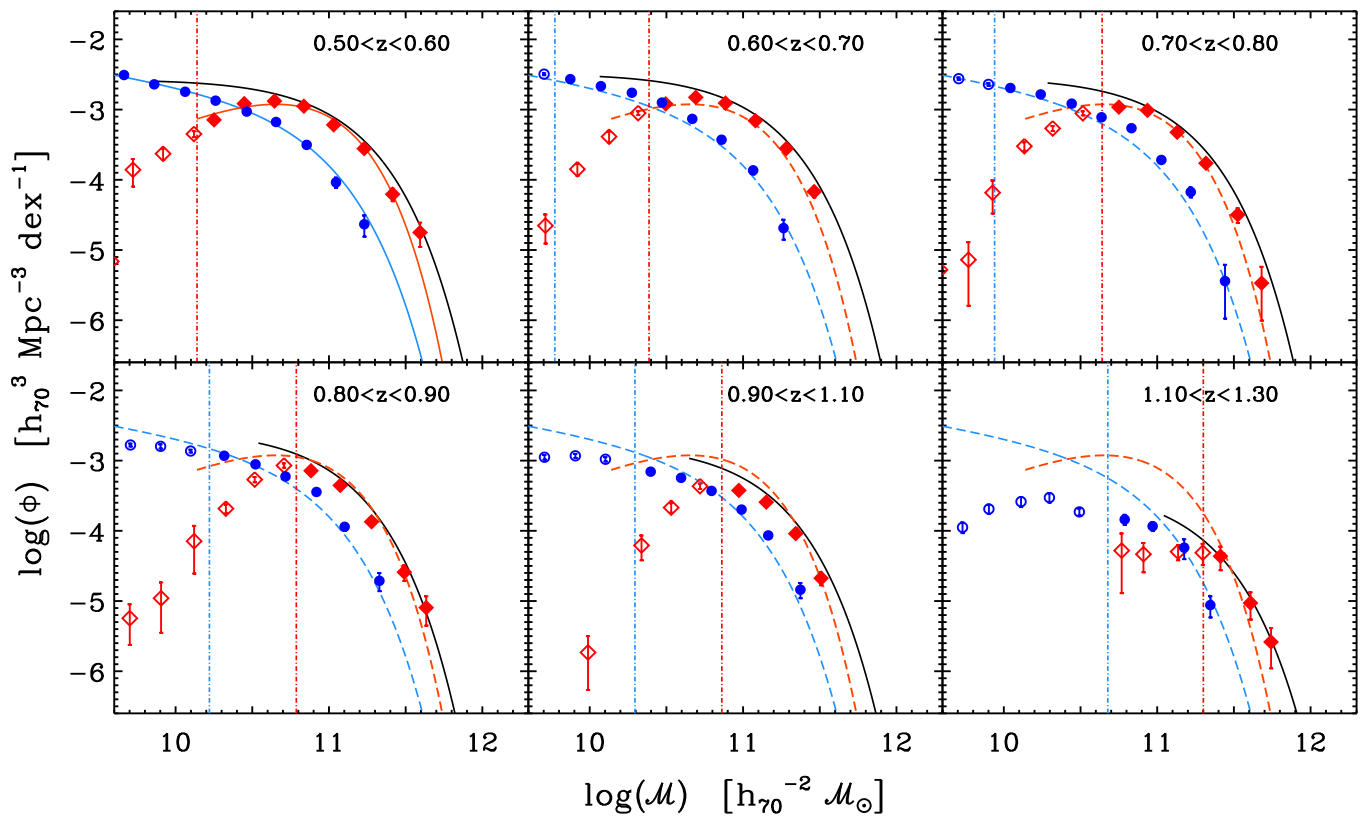}
  \caption{The galaxy stellar mass functions of the blue and red
    populations in VIPERS, derived using the $1/V_\mathrm{max}$.
    Symbols (circles and diamonds, respectively) are filled for data
    above the corresponding completeness limit
    $\mathcal{M}_\mathrm{lim}$ (vertical lines) and empty below. Error
    bars account for Poisson noise alone. The Schechter fit of the two
    populations in the bin $0.5<z<0.6$ (solid blue and red lines) is
    reported for reference as a dashed line in the other panels. The
    solid black line in each panel gives the Schechter best fit to the
    whole VIPERS sample in that redshift bin.}
  \label{blueredMF}
\end{figure*}

In order to distinguish the contribution of quiescent and actively
star forming galaxies to the global evolution, we now split the sample
according to the galaxy rest-frame $(U-V)$ colour (see \citealp{Fritz2013} for extensive discussion).


\subsection{Classification of galaxy types}
\label{Galaxy classification}
The absolute magnitudes for galaxies in the VIPERS catalogue were
computed from the same SED fitting procedure described in
Sect.~\ref{Stellar masses}, applying a k- and colour-correction,
derived from the best-fit SED, to the apparent magnitudes in the bands
that more closely match the rest-frame emission in the $U$ and $V$
filters \citep[see details in][]{Fritz2013}.  In this way, $(U-V)$
rest-frame colours can be reliably computed within the redshift range
of the survey, showing the classical bimodality and allowing us to
separate red-sequence from blue-cloud galaxies
\citep[cf.][]{Strateva2001,Hogg2002,Bell2004}.

The valley between the two populations is found to be slightly
evolving toward bluer colours at earlier epochs.  Despite its
simplicity, this photometric classification can be considered as a
good proxy for selecting quiescent and star-forming galaxies.  As
discussed by \citet{Mignoli2009} using zCOSMOS data, $86$\% ($93$\%)
of the galaxies selected as being photometrically red (blue) are also
quiescent (star-forming) according to their spectra.

To verify and validate our selection method, we also derived galaxy
photometric types by fitting our photometry with the empirical set of
$62$ templates used in \citet{Ilbert2006}, which was optimised to
refine the match between photometric and spectroscopic redshifts in
the VVDS.  The same set was also used to classify galaxies in several
other papers
\citep[e.g.][]{Zucca2006,Zucca2009,Pozzetti2010,Moresco2010}.  The
classification of VIPERS galaxies resulting from this second method
matches reasonably well with the $(U-V)$ colour selection.  More than
$70$\% of the red galaxies are defined as early-type objects by the
SED analysis, while more than $95$\% of blue galaxies are classifed as
late types.  For red galaxies this worsens beyond $z=1.1$, where only
$55$\% of the red galaxies are classified as early types in terms of
their SED.  In the same redshift range, instead, $98$\% of blue
galaxies are classified as late-type objects.

\subsection{Blue and red galaxy stellar mass functions}

Using this classification, we are now in a position to quantify the
contribution of red and blue galaxies to the GSMF and, in particular,
to its high-mass end. The results are shown in Fig.~\ref{blueredMF}.
The mass functions for each class are estimated in bins of $0.2$ in
$\log(\mathcal{M})$, using the same $1/V_\mathrm{max}$ method as
described in Sect.~\ref{Global GSMF}. Fits with the usual Schechter
function are provided, as described in the caption, to highlight
evolution (or absence thereof) as a function of redshift.

The predominance of red objects among the massive galaxies is clearly
visible in all redshift bins, with blue galaxies mainly contributing
at lower masses ($\mathcal{M}<\mathcal{M}_\star$).  Since the mass
completeness limit $\mathcal{M}_{\rm lim}$ for the blue population
extends to sufficiently low masses, we can perform the Schechter fit
by leaving $\mathcal{M}_\star$, $\Phi_\star$, and $\alpha$ free.  The
slope of the low-mass end remains almost constant in redshift for the
blue population, with $1.2<\alpha<1.3$, up to $z\simeq 0.9$, as seen
in previous works \citep[cf.][]{Pozzetti2010}.  At redshift higher
than this it can no longer be constrained.  With respect to the red
population, the high values of the mass completeness limit
$\mathcal{M}_\mathrm{lim}$ (see Sect.~\ref{Completeness}) prevent us
from studying the red sample in the same mass range; for instance, it
is not possible to determine the evolution of $\alpha$
\citep{Ilbert2010} or an upturn of the GSMF \citep[cf.][]{Drory2009}
in a reliable way.

From these measurements we can determine the value of
$\mathcal{M}_\mathrm{cross}$, where the blue and red GSMFs intersect,
i.e.~the dividing line between the ranges in which blue and red
galaxies respectively dominate the mass function
\citep{Kauffmann2003b}. The physical meaning of
$\mathcal{M}_\mathrm{cross}$ has been questioned \citep{Bell2007}, but
it is in general considered as a proxy to the transition mass of
physical processes such the quenching of star formation, (responsible
for the migration from the blue cloud to the red sequence), or the AGN
activity \citep[e.g.][]{Kauffmann2003c}. Moreover, its clear
dependence on environment \citep{Bolzonella2010} points to an
interpretation of the galaxy transformation that is not only linked to
secular processes.

\begin{figure}
\centering
  \includegraphics[width=0.48\textwidth]{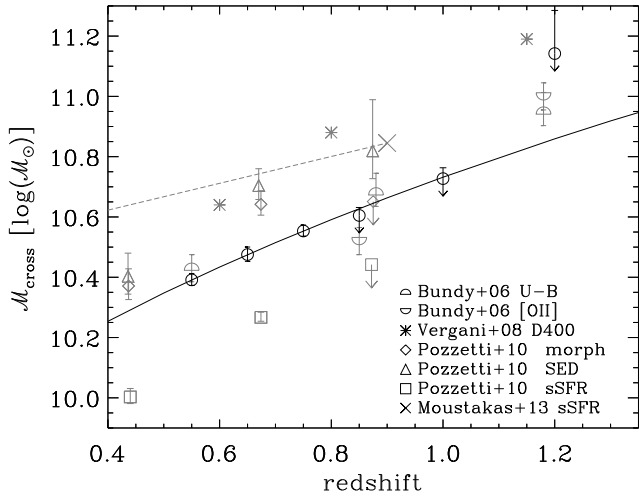}
  \caption{The values of the transition mass
    $\mathcal{M}_\mathrm{cross}$ as computed from
    Fig.~\ref{blueredMF}, plotted as a function of redshift.  The
    VIPERS measurements are given as black open circles, with a
    downward arrow when the transition mass is below the completeness
    mass of at least one of the two classes.  The solid line is a fit
    with a $(1+z)^3$ power law to the VIPERS points between $z=0.5$
    and $z=0.8$.  These are compared to literature estimates in grey.
    Points from \citet{Pozzetti2010} are obtained using three
    different classifications: a separation according to specific SFR
    (diamonds), a best-fit SED classification (triangles), and a
    morphological classification (squares). The points of
    \citet{Bundy2006} are based on either the $(U-B)$ bimodality or  
      [OII] emission (upper and lower half-circles respectively). 
      The points by \citet{Vergani2008}
    (asterisks) are based on a spectral classification (D4000 break).
    The value from PRIMUS \citep{Moustakas2013} at $z=0.9$ is reported
    as a cross, while the dashed line traces an evolution $\propto
    (1+z)^{1.5}$, as suggested in that paper; these authors classified
    active and quiescent galaxies with respect to their position in
    the SFR vs $\mathcal{M}$ diagram.}
  \label{Mcross}
\end{figure}

We quantify the value of the transition mass in each redshift bin
using the $1/V_\mathrm{max}$ measurements.  The transition mass
increases from $\log(\mathcal{M}_\mathrm{cross}/\mathcal{M}_\odot)=
10.4$ at $z\simeq0.55$ to
$\log(\mathcal{M}_\mathrm{cross}/\mathcal{M}_\odot) = 10.6$ at
$z\simeq0.75$, as shown in Fig.~\ref{Mcross}.  This trend is very well
fitted by a power law $\propto (1+z)^{3}$.  Beyond $z=0.8$ our
$\mathcal{M}_\mathrm{cross}$ estimates should be formally considered
as upper limits, since they fall below the mass completeness limit of
red galaxies, but at least up to $z=1.0$, they can be considered as a
good approximation of the real values, given their proximity to the
limit.
 
In Fig.~\ref{Mcross} we also plot results from previous studies.  In
this respect, it is important to underline that the value of
$\mathcal{M}_\mathrm{cross}$ provided by the various authors can
differ significantly from each other, depending on the adopted
classification.  For instance, the results of the morphological
classification used by \citet{Bundy2006} on the DEEP2 survey fall
above the mass ranges considered in the plot.  This could be related
to part of the ``red and dead'' galaxies at such redshifts becoming
ellipticals (in a morphological sense) at a later stage
\citep{Bundy2010}.  In fact, when we split the DEEP2 sample on the
basis of the $(U-B)$ bimodality, the results are in agreement with our
findings.  Our estimates of $\mathcal{M}_\mathrm{cross}$ are fairly
consistent (within $\pm0.2\,\mathrm{dex}$ approximatively) with those
of \citet{Vergani2008}, \citet{Pozzetti2010}, \citet{Moustakas2013}.
The estimates by \citet{Vergani2008}, also shown in Fig.~\ref{Mcross},
rely on the identification of the D$4000$ break in the VVDS spectra
and have a steeper redshift evolution,
$\mathcal{M}_\mathrm{cross}\propto (1+z)^{4}$.  \citet{Pozzetti2010}
derived $\mathcal{M}_\mathrm{cross}$ from the GSMFs of the zCOSMOS
(10k-bright) sample split using different criteria: a cut in specific
SFR (i.e.~sSFR $\equiv {\rm SFR}/\mathcal{M} \gtrless 10^{-1}
\mathrm{Gyr}^{-1}$), morphology (spheroidal vs disc/irregular
galaxies), and best-fit SEDs (same photometric types discussed in
Sect.~\ref{Galaxy classification}).  \citet{Moustakas2013} define
star-forming galaxies as lying in the so-called main sequence of the
SFR (estimated from the SED fitting) versus $\mathcal{M}$ diagram
\citep{Noeske2007}. They find a flatter evolution, with
$\mathcal{M}_\mathrm{cross} \propto (1+z)^{1.5}$.

%

\subsection{Evolution of the blue and red populations}

To collect further evidence of star-formation quenching processes that
cause the transition of galaxies from the so-called blue cloud to the
red sequence \citep{Faber2007}, we measured the evolution of the
galaxy number density of blue and red populations, namely $\rho_N^{\rm
  blue}(z)$ and $\rho_N^{\rm red}(z)$. These estimates are derived
using the $1/V_\mathrm{max}$ method, taking both Poisson noise and
cosmic variance into account.  We also verified, however, that the
results would essentially be the same if we had measured number
densities by integrating the Schechter best-fitting functions.  We
explore four narrow bins of stellar mass to highlight the dependence
of the quenching processes on this parameter. To improve statistics at
high stellar masses, we choose wider redshift bins here: $0.5$--$0.7$,
$0.7$--$0.9$, $0.9$--$1.1$, $1.1$--$1.3$.

\begin{figure*}
  \centering
  \includegraphics[width=0.9\textwidth]{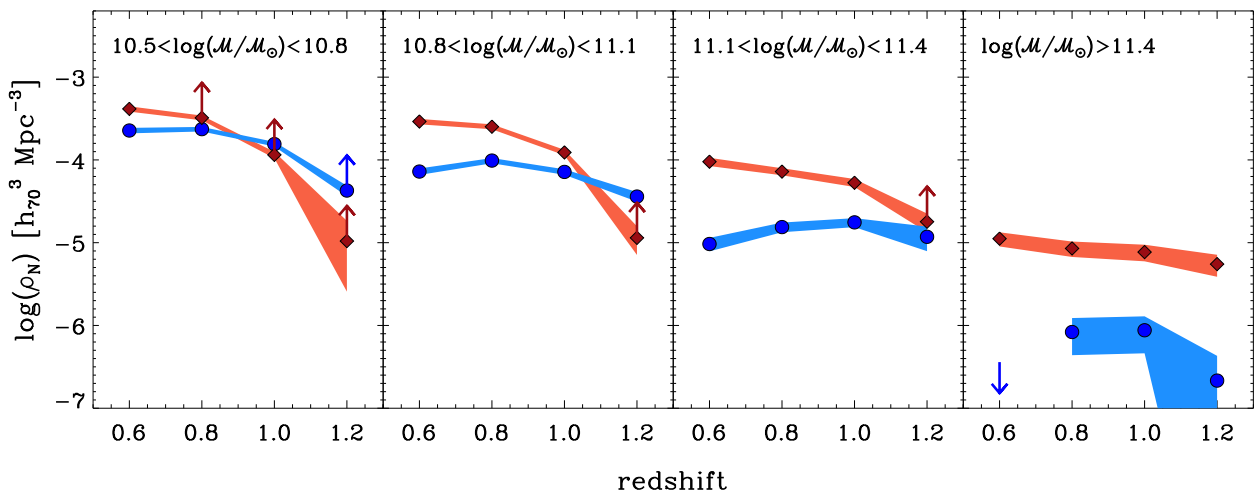}
  \caption{Evolution of the number density of the blue and red galaxy
    populations in VIPERS (filled circles and diamonds, respectively)
    with different stellar masses. Upward arrows represent lower
    limits when $\rho_N$ is estimated in a bin of mass affected by
    incompleteness, while a downward arrow represents the upper limit
    in case of zero detection (rightmost panel).  The error corridors
    reflect the overall uncertainties, which include both Poisson
    noise and cosmic variance added in quadrature.}
  \label{blueredND}
\end{figure*}

At intermediate masses
($10.8<\log(\mathcal{M}/\mathcal{M}_\odot)<11.1$), the number density
of red galaxies $\rho_N^{\rm red}$ increases by a factor of $\sim2.5$
from $z\!=\!1$ to $z\!=\!0.6$, whereas at higher masses the variation
is much smaller. Red galaxies with mass $11.1 <
\log(\mathcal{M}/\mathcal{M}_\odot) < 11.4$ evolve from a number
density $\rho_N^{\rm red} = (5.3 \pm 0.6) \times
10^{-5}\,\mathrm{Mpc}^{-3}\,h_{70}^3$ to $(9.5 \pm 1.1)\times
10^{-5}\,\mathrm{Mpc}^{-3}\,h_{70}^3$ in the same redshift interval
(about $80$\% increase).  The increase is even smaller ($45\%$) for
galaxies with $\log(\mathcal{M/M}_\odot) \geqslant 11.4$.  With the
VIPERS data we are able for the first time to provide significant
evidence of this trend for such massive galaxies at these redshifts.
This result is in line with the mass-assembly downsizing scenario
highlighted in previous works
\citep{Cimatti2006,Pozzetti2010,Ilbert2010}: barring systematic
effects due to the uncertainty on $\mathcal{M}$, red galaxies with
$\mathcal{M} > 10^{11}\,\mathcal{M}_\odot$ build their stellar mass
well before the less massive ones and do not experience any strong
evolution between $z\simeq 1.2$ and $z \simeq 0.6$.  At these
redshifts, quenching mechanisms seem to be more efficient at low and
intermediate masses, as also recently suggested by
\citet[][]{Moustakas2013}.  With respect to PRIMUS, the VIPERS survey
extends this finding to higher masses ($\log(\mathcal{M/M}_\odot) >
11.4$) and redshifts (up to $\simeq 1.2$).  The evidence of mass
dependence of quenching agrees, for instance, with
\citet[][]{Peng2010}, although other mechanisms could play a
non-negligible role \citep[e.g.~galaxy mergers,][]{Xu2012}.

The co-moving number density of blue galaxies is instead found to be
relatively stable between $z\simeq 1$ to $z\simeq 0.6$ for objects
with mass $10.5 \leqslant \log(\mathcal{M/M}_\odot) < 10.8$, with a
$10\%$ variation.  For higher mass blue galaxies, for which the sample
is complete at all redshifts, the density $\rho_N^\mathrm{blue}$ of
objects with mass $10.8 \leqslant \log(\mathcal{M/M}_\odot) <11.1$
indicates a mild increase between $z\simeq 1.2$ and $z=0.8$.

The most massive blue galaxies ($\log(\mathcal{M/M}_\odot)\geqslant
11.4$) disappear at $z\lesssim0.6$ (see the right panel in
Fig.~\ref{blueredND}), suggesting that, at such high masses, star
formation already turns off at earlier epochs (i.e.  $z>1.3$).  When
the whole VIPERS sample is available, we will continue the analysis of
the massive-end build-up with more robust statistics.  Moreover, a
step forward for a better comprehension of this picture will be the
use of spectral features to determine reliable estimates of the SFR,
and therefore to better separate passive and active galaxies.

\section{Conclusions}
\label{Conclusions}

We measured the GSMF between $z=0.5$ and $z=1.3$ using the first data
release of VIPERS.  The forthcoming VIPERS Public Data Release 1
(PDR-1) will contain the catalogue of the $53\,608$ spectroscopic
galaxy redshifts used in the present analysis.  The galaxy stellar
masses were estimated through the SED fitting technique, relying on a
large photometric baseline and, in particular, on a nearly full
coverage of our fields with near-infrared data.  We performed several
tests to verify that the systematics intrinsic to the method of SED
fitting (e.g.~the parametrisation of the SFH) do not introduce any
significant bias into our analysis. The large volume probed by VIPERS
results in extremely high statistics, dramatically reducing the
uncertainties due to Poisson noise ($\sigma_\mathrm{Poiss}$) and
sample variance ($\sigma_\mathrm{cv}$).  We estimated the latter by
using $57$ galaxy mock catalogues based on the MultiDark simulation
\citep{Prada2012} and the stellar-to-halo mass relation of
\citet{Moster2013}. These mocks closely reproduce the characteristics
of the VIPERS survey.

We empirically determined a completeness threshold
$\mathcal{M}_\mathrm{lim}$ above which the mass function can be
considered complete. This limiting mass evolves as a function of $z$,
ranging from $\log(\mathcal{M/M}_\odot)=9.8$ to $11$ in the redshift
interval $0.5$--$1.1$.  We focussed our analysis on the high-mass end
of the GSMF, where VIPERS detects a particularly high number of rare
massive galaxies.  The main results we obtain follow.

\begin{itemize}

\item VIPERS data tightly constrain the exponential tail of the
  Schechter function, which does not show significant evolution at
  high masses below $z=1.1$. The same result is provided by analysis
  of the co-moving number density $\rho_N$, calculated in different
  bins of stellar mass. At $z \simeq 1.2$ most of the massive galaxies
  with $\log(\mathcal{M/M}_\odot) \geqslant 11.4$ are already in
  place, whereas below $\log(\mathcal{M/M}_\odot)= 11.4,$ the galaxy
  number density increases by a factor of $\sim 3.5$ from $z \simeq
  1.2$ to $z \simeq 0.6$.

\item We compared our observed GSMFs with those derived from
  semi-analytical models
  \citep{DeLucia&Blaizot2007,Bower2006,Guo2011}.  While the
  discrepancy at low masses between models and observations is well
  established and has been exhaustively discussed in literature,
  predictions at the high-mass end of the GSMF have not yet been
  verified with sufficient precision. We show that the high accuracy
  of the VIPERS mass functions makes them suitable for this kind of
  test, although further improvement to reduce stellar mass
  uncertainties would be beneficial.  From a first analysis, the
  VIPERS data appear to be consistent with the \citet{Guo2011} model
  at $\log(\mathcal{M/M_\odot})\geqslant11$, once the uncertainties in
  the stellar mass estimates are taken into account.  A more detailed
  analysis will be the subject of a future work.  We suggest that
  VIPERS GSMFs can be effectively used to constrain models at multiple
  redshifts simultaneously, in small steps of $\Delta z$.  This could
  shed light on the time scale of the physical mechanisms that
  determine the evolution at higher masses (for instance, the
  AGN-feedback efficiency).

\item We divided the VIPERS sample by means of a colour criterion
  based on the $(U-V)$ bimodality \citep{Fritz2013} and estimated
  the blue and red GSMF in the same range, $0.5 < z < 1.3$.  We find
  that the transition mass above which the GSMF is dominated by red
  galaxies is about
  $\log(\mathcal{M_\mathrm{cross}/M}_\odot)\simeq10.4$ at $z \simeq
  0.55$ and evolves proportional to $(1+z)^3$.

\item The number density of the red sample shows an evolution that
  depends on stellar mass, being steeper at lower masses.  At high
  stellar masses, the quenching of active galaxies has not been
  thoroughly studied because of their rareness.  We obtained a first
  impressive result with VIPERS, by detecting at $z\simeq 1$ a
  significant number of very massive active galaxies with
  $\log(\mathcal{M/M}_\odot)\geqslant 11.4$, which have all migrated
  onto the red sequence by $z=0.6$, i.e.~in about $2$\,Gyrs.

\end{itemize}

The first data release of VIPERS has allowed us to study the evolution
of the galaxy stellar mass function over an unprecedented volume at
redshifts $z=0.5-1.3$.  We emphasise the constraining power of this
dataset, particularly for the abundance of the most massive galaxies,
both quiescent and star-forming.  In forthcoming studies we will make
full use of the growing sample and of the measurement of spectral
features, in order to investigate the cosmic star formation history
and compare galaxy formation models at high redshift.

\begin{acknowledgements}
  We are grateful to Lucia Pozzetti for useful suggestions and for
  providing zCOSMOS results. We thanks Simon J.~Mutch who provided the
  GSMF foreseen by the model described in \citet{Mutch2013} in our
  preferred redshift bins.  ID warmly thanks Ivan Delvecchio for
  useful discussions.

  We acknowledge the crucial contribution of the ESO staff for the
  management of service observations. In particular, we are deeply
  grateful to M. Hilker for his constant help and support of this
  programme.  Italian participation in VIPERS has been funded by INAF
  through PRIN 2008 and 2010 programmes. LG and BRG acknowledge
  support of the European Research Council through the Darklight ERC
  Advanced Research Grant (\# 291521). OLF acknowledges support of the
  European Research Council through the EARLY ERC Advanced Research
  Grant (\# 268107).  Polish participants have been supported by the
  Polish Ministry of Science (grant N N203 51 29 38), the Polish-Swiss
  Astro Project (co-financed by a grant from Switzerland, through the
  Swiss Contribution to the enlarged European Union), the European
  Associated Laboratory Astrophysics Poland-France HECOLS and a Japan
  Society for the Promotion of Science (JSPS) Postdoctoral Fellowship
  for Foreign Researchers (P11802). GDL acknowledges financial support
  from the European Research Council under the European Community's
  Seventh Framework Programme (FP7/2007-2013)/ERC grant agreement n.
  202781. WJP and RT acknowledge financial support from the European
  Research Council under the European Community's Seventh Framework
  Programme (FP7/2007-2013)/ERC grant agreement n. 202686. WJP is also
  grateful for support from the UK Science and Technology Facilities
  Council through the grant ST/I001204/1. EB, FM and LM acknowledge
  the support from grants ASI-INAF I/023/12/0 and PRIN MIUR 2010-2011.
  YM acknowledges support from CNRS/INSU (Institut National des
  Sciences de l'Univers) and the Programme National Galaxies et
  Cosmologie (PNCG). CM is grateful for support from specific project
  funding of the {\it Institut Universitaire de France} and the LABEX
  OCEVU.
\end{acknowledgements}

\bibliography{gsmf_aa}

\end{document}